\documentclass[%
 reprint,
superscriptaddress,
preprintnumbers,
 amsmath,amssymb,
 aps,
]{revtex4-2}

\usepackage{graphicx}
\usepackage{dcolumn}
\usepackage{bm}
\usepackage{simplewick}

\begin{document}


\title{The Strong Field QED approach of the vacuum interaction processes at
ELI-NP}%

\author{M. Pentia} 
\thanks{Corresponding author: pentia@nipne.ro}
\affiliation{IFIN-HH, National Institute for Physics and Nuclear Engineering, Reactorului 30,
RO-077125, POB-MG6 Bucharest-Magurele, Romania}
\author{C.R. Badita}
\affiliation{IFIN-HH, National Institute for Physics and Nuclear Engineering, Reactorului 30,
RO-077125, POB-MG6 Bucharest-Magurele, Romania}
\author{D. Dumitriu}
\affiliation{IFIN-HH, National Institute for Physics and Nuclear Engineering, Reactorului 30,
RO-077125, POB-MG6 Bucharest-Magurele, Romania}
\author{A.R. Ionescu}
\affiliation{IFIN-HH, National Institute for Physics and Nuclear Engineering, Reactorului 30,
RO-077125, POB-MG6 Bucharest-Magurele, Romania}
\author{H. Petrascu}
\affiliation{IFIN-HH, National Institute for Physics and Nuclear Engineering, Reactorului 30,
RO-077125, POB-MG6 Bucharest-Magurele, Romania}




\date{\today}

\begin{abstract}
The commissioning of the high power laser facility Extreme Light 
Infrastructure - Nuclear Physics (ELI-NP) at Bucharest-M\u agurele 
(Romania) allows the in-depth study of nonlinear 
interactions in Strong Field Quantum Electrodynamics (SF-QED). 
The present paper analyzes the SF-QED processes possible to study at ELI-NP. 
Carrying out such experiments will allow finding answers to many fundamental 
QED questions. Firstly it needs to highlight and evaluate the various 
interactions with the virtual particles of the QED vacuum as the 
processes of multi-photon inverse Compton scattering, $e^+e^-$ pair 
production, $e^+e^-$ pair annihilation, $e^-e^-$ Moller scattering, $e^+e^-$ 
Bhabha scattering, electron self-energy, photon self-energy or vacuum 
energy. In this sense, the current worldwide results are analyzed along with 
the main steps necessary for the design of SF-QED experiments at ELI-NP. 

After a brief review of the first experiment (E-144 SLAC) which confirmed 
the existence of nonlinear QED interactions of high-energy electrons with 
photons of a laser beam, we presented the fundamental QED 
processes that can be studied at ELI-NP in the multi-photon regime along 
with the characteristic parameters of the laser beam used 
in the QED interaction with electrons.

To prepare an experiment at ELI-NP, it is necessary to analyze both the kinematics 
and the dynamics of the interactions. Therefore, we first 
reviewed the kinematics of linear QED processes and then the corresponding 
Feynman diagrams. For nonlinear, non-perturbative multi-photon QED 
interactions, the Feynman diagram technique must be adapted from linear 
to nonlinear processes. This is done by switching to quantum fields 
described by Dirac-Volkov dressed states of particles in an intense 
electromagnetic (EM) field. This allows the evaluation of the 
amplitude of the physical processes and finally the determination of 
the cross-sections of these processes.

SF-QED processes of multi-photon interactions with strong laser fields 
can be investigated taking into account the characteristics of the 
ELI-NP facility in the context of QED vacuum pair production of 
electron-positron pairs and energetic gamma rays.

Finally, we present some similar experimental projects from other 
research centers, in different stages of implementation.

\end{abstract}

    

\maketitle


\section{Introduction}
\label{sec:intro}

In 2009 G.V. Dunne (University of Connecticut) \cite{DUNE09} remarked 
"the ELI project opens up an entirely new non-perturbative 
regime of QED and of quantum field theories in general. 
There are many experimental and theoretical challenges ahead. Theoretically, 
the biggest challenge in the non-perturbative arena is to develop efficient 
techniques, both analytical and numerical, for computing the effective 
action and related quantities, in external fields that realistically 
represent the experimental laser configurations. A lot of progress has 
been made in this direction, but new ideas and methods are still needed".

Most of the high power laser works interpret the SF-QED interactions as 
non-perturbative processes, relative to the clasical field-matter interaction 
strength $\xi$ and quantum parameter $\chi$ (see later). 
For experimental stu\-dies must be evaluated the cross sections of 
these processes.
Therefore, the treatment of SF-QED processes must be done in terms of
vacuum interaction processes with Feynman diagrams for "dressed" particle
in oscillatory EM fields \cite{VOLK35}.
The treatment of non-perturbative QED has a different meaning than that 
of non-perturbative QCD. The last one is connected with the strong coupling 
parameter $\alpha_s$, which is too large for a perturbative approach. 
On the other hand, the non-perturbative QED regime is characterized 
by the coupling parameter $\xi$ between the matter and 
the laser field. Here we are dealing with SF-QED interactions but we do 
not appeal to the QED coupling constant $\alpha\approx 1/137$ as 
parameter of perturbative developments, but to the coupling $\xi$ 
between matter and the laser field, which can exceed the unit. 
As such, processes and diagrams of given order in $\alpha$ (Feynman diagrams) 
depend on all terms of the expansion in $\xi$.
On the other hand, the electron-laser interactions are described by particle 
"dressed" states \cite{RITU85} as it comes to have multi-photon interactions, and by 
high-order processes with radiative corrections in the theory.

Today, the ELI-NP facility can provide laser beams of 2 x 10 PW with 
intensities up to $10^{22} - 10^{23}~ W/cm^2$
\cite{TURC19,NEGO16,GALE18}. Therefore, 
we can proceed to initiate experimental works for the in-depth study of 
nonlinear QED processes. With such laser beams can be performed a series 
of works as:

\begin{itemize}
        \item
Systematic studies of the dynamics of fundamental QED processes possible 
to approach at high power lasers and to evaluate the amplitude of these
processes such as: $\gamma e$ inverse Compton scattering 
\cite{KIBB65,BERE82,CHEN09,PIKE14a,ROSE20}, Breit-Wheeler 
$e^+e^-$ pair production \cite{BREI34}, Bethe-Heitler $e^+e^-$ pair production 
\cite{BETH34}, Dirac $e^+e^-$ pair annihilation, $e^-e^-$ Moller Scattering, 
$e^+e^-$ Bhabha Scattering, Electron Self Energy, Photon Self Energy, 
Vacuum Energy \cite{PESK95}.
        \item
Proposal of experimental works for the measurement of physical properties 
related to the production of $e^+e^-$ pairs (Schwinger mechanism) in the 
photon-multi-photon interaction (Breit-Wheeler nonlinear), the 
multi-photon-virtual photon interaction of the nucleus field 
(Bethe-Heitler nonlinear), or the production and measurement 
of QED (positronium) bound states.
        \item
Design and carry out experimental works to measure some fundamental 
processes using high-power lasers at ELI-NP.
\end{itemize}

The production of a large number of positrons with 
MeV energies opens the door to new avenues of antimatter research, 
including understanding the physics of various processes and phenomena 
in astrophysics, such as black holes and gamma-ray bursts 
\cite{WARD98,MESZ02}, or from pair plasma physics \cite{WELD91,BLAC93}.

\section{Fundamental gamma-electron interactions}

The conversion of laser light to matter is one of the fundamental processes 
of EM photon-photon interaction, less experimentally 
studied so far. This is one of the spectacular predictions of QED, but 
difficult to achieve. The reason is due both to a reduced photon-photon 
cross section ($\sim$ 0.1 barn), but mainly due to the difficulty 
of achieving an adequate density of photon beams \cite{BREI34}. 

However, E.J. Williams \cite{WILL34} noted that sufficiently 
high photon densities can be obtained in the intense electric field of a 
nucleus in relativistic motion. Therefore, the production of pairs of 
particles from the EM field is possible to achieve in principle:
\begin{itemize}
        \item
In a static electric field - the Schwinger effect \cite{SAUT31};
        \item
In a photonic field - Breit-Wheeler production \cite{BREI34,REIS62,NARO65};
        \item
In a combination of the two - Bethe-Heitler production \cite{BETH34}.
\end{itemize}

\begin{figure}[h]
\hspace{-5mm}
\includegraphics[width=90mm]{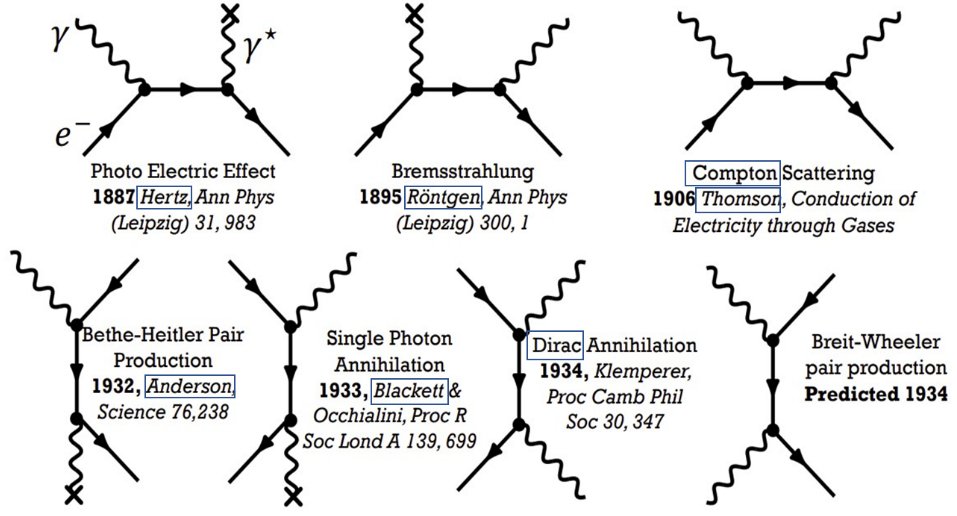}
\vspace{-5mm}
\caption{
Light ($\gamma$) and matter ($e^-$) QED interaction processes. 
(thanks to Oliver Pick presentation 23 October 2014, LNFN – Frascati, 
"Observing the two-photon Breit-Wheeler process for the first time", 
underlined - Nobel laureates)} 
\label{fig:qed-processes}
\end{figure}

Such linear QED laser light - matter interaction processes are shown in 
Fig.\ref{fig:qed-processes} in the form of Feynman diagrams. They can be 
studied at the ELI-NP facility, but in the laser multi-photon oscillatory EM field, 
as nonlinear interactions of "dressed" particles \cite{VOLK35}.

\begin{figure}[ht]
\hspace{20mm}
\begin{tabular}{p{40mm}p{40mm}}
\begin{center}
\vspace{-4mm}
\includegraphics[width=30mm]{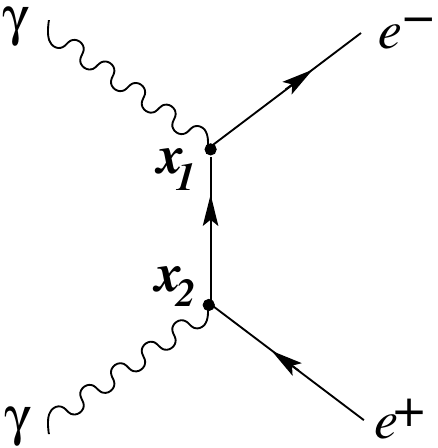}
\end{center}
&
\begin{center}
\includegraphics[width=40mm,height=22mm]{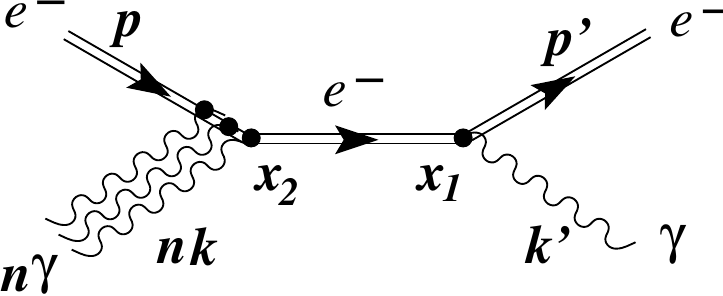}
\end{center}
\\[-33pt]
\begin{center}
(a)
\end{center}
&
\begin{center}
(b)
\end{center}
\\[-21pt]
\begin{center}
\vspace{-2mm}
\includegraphics[width=30mm]{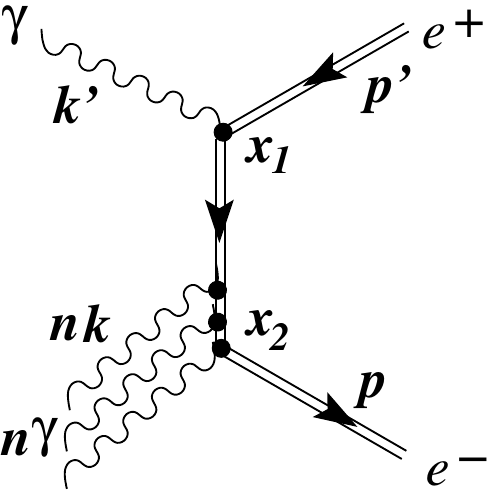}
\end{center}
&
\begin{center}
\includegraphics[width=30mm]{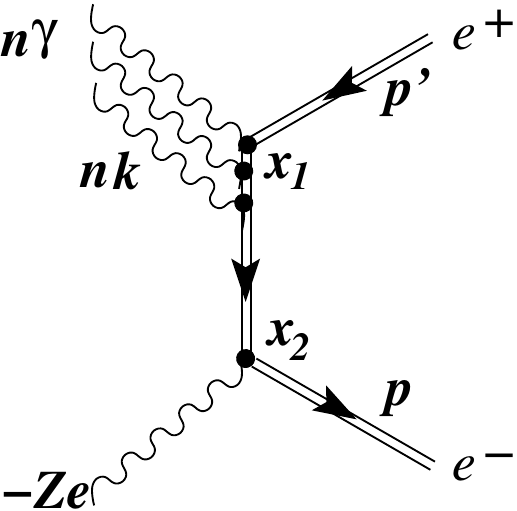}
\end{center}
\\[-33pt]
\begin{center}
(c)
\end{center}
&
\begin{center}
(d)
\end{center}
\end{tabular}
\vspace{-2mm}
\caption{
a) Breit-Wheeler linear process; b) Nonlinear inverse Compton scattering 
(multi-photon); c) Nonlinear Breit-Wheeler process (multi-photon); 
d) Nonlinear Bethe-Heitler interaction.
}
\label{fig:proc-Feynman}
\end{figure}

Current theoretical and experimental works 
with high-power lasers (see for example \cite{ILDE23}) have 
highlighted the possibility of experimental study of some 
fundamental QED interactions such as:
\begin{itemize}
\item
 Linear Breit-Wheeler interaction process (Fig.\ref{fig:proc-Feynman}.a): \\ 
$\gamma + \gamma \to e^+ + e^-$ with 
electron-positron pair production, treated in linear QED \cite{GREI08}.
\item
 Nonlinear inverse Compton scattering (multi-photon) 
(Fig.\ref{fig:proc-Feynman}.b): \\ 
$e^- + n\,\gamma_L \to e^- + \gamma$, where the initial electron interacts 
with $n$ laser photons $\gamma_L$ and is emitted a $\gamma$ photon of high 
energy. Up to 40\% of the energy of the initial electrons is transferred to 
the final photon \cite{BULA96}.
\item
 Nonlinear (multi-photon) Breit-Wheeler interaction process 
(Fig.\ref{fig:proc-Feynman}.c): \\
$\gamma + n\,\gamma_L \to e^+ + e^-$, pair production, with the energy 
of several laser photons transformed into the mass of electron-positron 
pairs \cite{E-144}.
\item
 The Bethe-Heitler interaction process with the intense electric field 
of the nucleus (Fig.\ref{fig:proc-Feynman}.d):\\ 
$n \gamma_L + \gamma_V \to e^+ + e^-$, where the $n \gamma_L$ the laser 
multi-photon interaction with $\gamma_V$ virtual photon of the intense field 
of the nucleus, leads to $e^+e^-$ pair production \cite{BETH34}.
\end{itemize}

\subsection{QED vacuum interaction processes}

\begin{figure}[ht]
\hspace{20mm}
\begin{tabular}{p{40mm}p{40mm}}
\begin{center}
\vspace{-2mm}
\includegraphics[width=35mm]{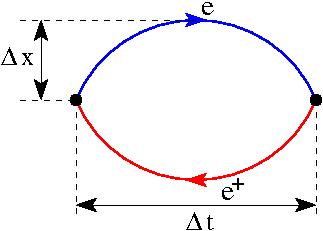}
\end{center}
&
\begin{center}
\includegraphics[width=35mm]{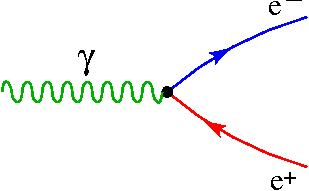}
\end{center}
\\[-27pt]
\begin{center}
(a)
\end{center}
&
\begin{center}
(b)
\end{center}
\end{tabular}
\caption{
a) Vacuum fluctuations with creation and annihilation of virtual $e^+e^-$ 
pairs; b) Energy transfer from the EM field ensures the virtual pairs 
separation and their transformation into real pairs.
}
\label{fig:vacuum-fluct}
\end{figure}

Under normal conditions, the physical vacuum, due to quantum fluctuations, 
is in a permanent state of "boiling", with the creation and annihilation of 
virtual particle-antiparticle pairs. According to the Heisenberg principle, 
locally, on short time intervals $\Delta t$, there are energy fluctuations 
$\Delta E$, so that their product is not smaller than $\hbar$: \quad 
$\Delta E \cdot \Delta t \ge\hbar$. \quad
On time intervals $\Delta t$, the $\Delta E$ fluctuation can produce 
$e^+e^-$ pairs, at shallow depth under the mass shell. That is, 
$\Delta E \approx 2m_ec^2 = 2 \cdot 0.511 MeV \approx 10^6$ eV and 
$\Delta t \ge \hbar / 2m_e\, c^2 = 10^{-22}$ s.
So, locally, the process of $e^+e^-$ pair production is confined in a 
spatial interval $\Delta x$ and temporal $\Delta t$, after which the 
pair annihilates (see Fig.\ref{fig:vacuum-fluct}.a)). 
This process of production and annihilation of virtual 
$e^+e^-$ pairs is associated with the vacuum polarization phenomenon.

If an external electric field $E$ transfers enough energy to these virtual 
pairs during $\Delta t$ (see Fig.\ref{fig:vacuum-fluct}.b)) it can transform 
them into real pairs that can be observed and recorded experimentally.
 
The characteristic distance $2\Delta x$ over which the electric field can 
produce real $e^+e^-$ pairs is given by the reduced Compton wavelength 
$\lambdabar_c$,
\begin{equation}\label{eq:compt-wavelength} 
2\Delta x \approx 2\, c\,\Delta t = \frac{\hbar}{m_e c}
=\lambdabar_c \approx 386\cdot 10^{-15} m
\end{equation}

The real $e^+e^-$ pair will be produced if a minimum energy W is transferred 
from the field E,
\begin{equation}\label{eq:min-energy} 
W=e\,E\, \lambdabar_c =\frac{eE\hbar}{m_e\, c} > 2 m_e\, c^2
\end{equation}

Hence the minimum value of the electric field leading to vacuum breakdown is:

\begin{equation}\label{eq:min-el-field} 
E> \frac{2m_e^2\, c^3}{\hbar e}=2E_{cr}
\end{equation}

In the case of a uniform EM field, this defines the Schwinger limit of the 
critical electric field $E_{cr}$ capable of producing vacuum breakdown, i.e. 
the starting value of the spontaneous real $e^+e^-$ pairs production in the 
laser field - vacuum interaction:
\begin{equation}\label{eq:Schwingwr-field} 
E_{cr} = \frac{m_e^2 c^3}{\hbar e}=1.323\cdot 10^{18}\, V/m
\end{equation}

Other words, the value (\ref{eq:Schwingwr-field}) specifies the 
minimum electric field capable of performing on an electron the work 
$\epsilon_e=W/2$ over the Compton wavelength\,\, $\lambdabar_c$, 
equal to its rest mass $m_e\, c^2$: 
\begin{equation}\label{work-el-mass} 
\epsilon_e = e\, E_{cr}\, \lambdabar_c= m_e\, c^2 
\end{equation}

The value (\ref{eq:Schwingwr-field}) of the 
Schwinger critical electric field $E_{cr}$ has not 
yet been reached experimentally. This field leading to vacuum breakdown 
with the production of $e^+e^-$ pairs  (Breit-Wheeler process), 
presents a lot of nonlinear aspects.

The field intensities accessible in the laboratory system nowadays are 
three orders of magnitude lower than the value of the critical field $E_{cr}$. 
However, in the rest system of a high energy electron, it "sees" the 
transverse component of the laser electric field $E$ boosted by $\gamma_e$ 
(Lorentz factor) and reaches a value $E^* = \gamma_e\, E$.

Based on the relationship between the electric field $E$ and the intensity 
of the laser beam $I_L$: $E(V/m) = 2750\sqrt{I_L}\, (W/cm^2)$,\,\, for the ELI-NP 
with the intensity $I_L\!>\! 10^{22} W/cm^2$, light pulses lead 
to field intensity $E \approx 10^{14}\, V/m$ at the focal point, on a distance 
of seve\-ral lasr wavelengths. For example, with an electron beam 
energy of $\epsilon_e = 10$ GeV, the Lorenz factor is 
$\gamma_e = \epsilon_e/m_ec^2 \approx 2\cdot 10^4$ so that if a laser beam 
collides head-on, it "sees" a boosted laser field $E^* \approx 2 \cdot 10^{18}$ 
V/m. Therefore, the intensity of the laser field in the moving electron 
system will be of the order of critical value $E_{cr}$ (\ref{eq:Schwingwr-field}).

\subsection{First matter-light conversion experiment (E-144 SLAC)}

\begin{figure}[h]
\hspace{-2mm}
\begin{minipage}{85mm}
\includegraphics[width=90mm]{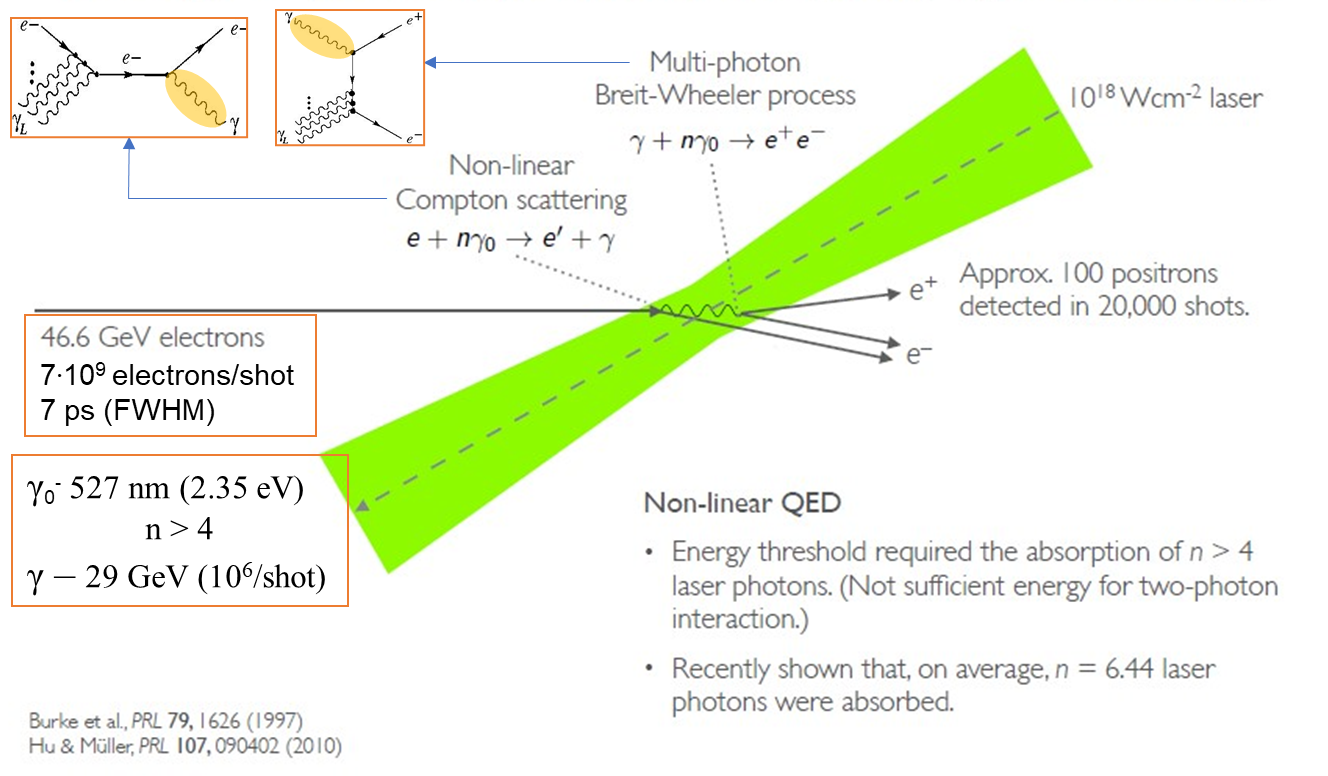}
\caption{
The SLAC E-144 experiment. Positron production process in 
$\gamma\gamma$  scattering \newline
{\scriptsize (O.Pike, https://agenda.infn.it/event/8532/contributions/74190)}
} 
\label{fig:SLAC-E-144}
\end{minipage}
\end{figure}

In the SLAC E-144 experiment \cite{BULA96,E-144,BURK97,BAMB99}, 
studies of nonlinear QED processes were carried out using 46.6 GeV 
electrons scattered on the intense laser EM field with a wavelength of 
527 nm (2.35 eV), see Fig.\ref{fig:SLAC-E-144}. Using peak 
focused laser in pulses of the order of terawatt and energy of 650 mJ, 
intensities of $10^{18}~W/cm^2$ were obtained.

The production of $e^+e^-$ pairs requires an energy in the center of mass 
system at least of $2 m_ec^2 = 1.02$ MeV. This can be achieved by a 
nonlinear Breit-Wheeler process, i.e. by scattering on a laser beam a 
high-energy photon created by inverse Compton scattering of a laser beam 
on a high-energy electron (see Fig.\ref{fig:SLAC-E-144}). 

For the production of pairs from the interaction 
$\gamma_1\gamma_2 \to e^+e^-$, having laser light $\gamma_1$ of wavelength 
527 nm (energy 2.35 eV), would require photons $\gamma_2$ of energy 
109 GeV. But with 527 nm laser photons scattered on a 46.6 GeV electron 
beam available at SLAC, the maximum energy of the nonlinear Compton 
scattered photons is only 29.2 GeV \cite{BAMB99}.
Therefore it is necessary to use both the relativistic electron boost and 
a multi-photon interaction process. The electric field increase is obtained 
by the relativistic boost of the counter-propagating electron with the 
energy in the laboratory system $\epsilon_e$ by a Lorentz factor 
$\gamma_e\!=\!\epsilon_e/m_e\,c^2\!\gg\!1$. Then the electron "sees" the $E_L$ 
laser field boosted to the value $E^*\!=\!\gamma_e E_L$. 
So, for $\epsilon_e\!=\!46.6$~GeV, the electron with Lorentz 
factor $\gamma_e\!=\!46.6 \cdot 10^3$~MeV$ / 0.511~MeV\!=\!9 \cdot 10^4$ 
colliding head-on with a 527 nm laser photon, "sees" the field in its 
rest frame $E^* = 1.1 \cdot 10^{18}$~V/m. This is close to the Schwinger 
limit $E_{cr}= m_e^2\,c^3/\hbar\,e =1.3\cdot 10^{18}$~V/m, which can 
transfer to an electron, along a Compton wavelength, the energy equal to its 
rest mass, at which a static electric field would vacuum spontaneously 
break-down into $e^+e^-$ pairs. 

To reach the value of the critical field $E_{cr}$, we must resort to 
multi-photon and electron interactions with a high-intensity laser $I_L$, 
which in this way ensures a sufficiently strong electric field $E_L$, 
according to $I_L\!=\!\epsilon_0\,c\,\big<E_L^2\big>$ 
(see later) (\ref{laser-intensity}).

\begin{figure}[h]
\centering
\begin{tabular}{m{40mm}m{40mm}}
\begin{minipage}{40mm}
\includegraphics[width=40mm]{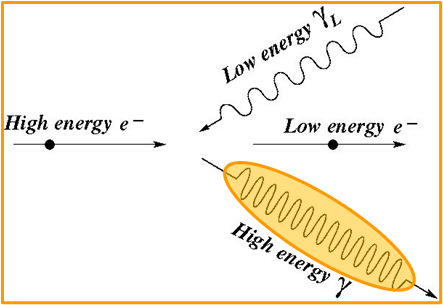}
\end{minipage}
&
\begin{minipage}{40mm}
\includegraphics[width=40mm]{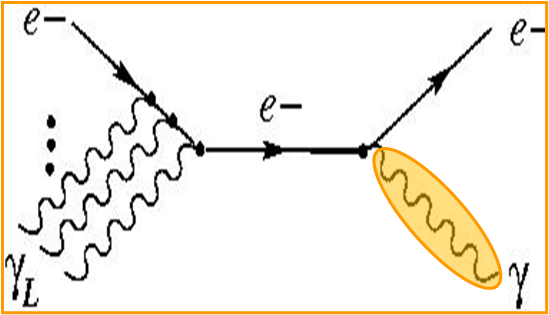}
\end{minipage}
\end{tabular}
\caption{
High-energy photon production by inverse Compton scattering of 
laser beams.
\newline
a) Kinematics of the Compton inverse scattering; b) Feynman diagram for 
determining the amplitude.
} 
\label{fig:inv-Compt-scatt}
\end{figure}

\begin{figure}[h]
\hspace{12mm}
\begin{tabular}{m{45mm}m{45mm}}
\centering
\includegraphics[width=35mm]{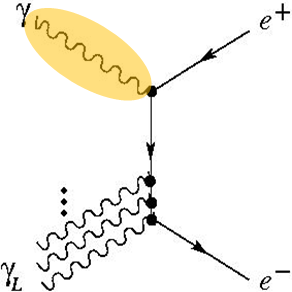}
&
\includegraphics[width=35mm]{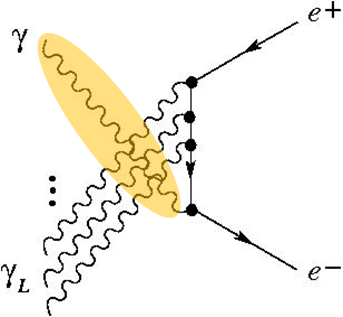}
\end{tabular}
\caption{
Feynman diagrams for the production of $e^+e^-$ pairs by the nonlinear 
Breit-Wheeler process.
} 
\label{fig:BW-nonlinear}
\end{figure}

The parameters of the experiment correspond to the nonlinear Compton regime 
(see Fig.\ref{fig:inv-Compt-scatt}). 
The high-energy photon resulting from inverse Compton scattering interacts 
with the multi-photon laser beam and produces $e^+e^-$ pairs through the 
nonlinear Breit-Wheeler process (see Fig.\ref{fig:BW-nonlinear}).

 \vspace{3mm}

{\bf Important Note:}
Linear, single-photon QED interaction processes described by Feynman 
diagrams such as those in Fig.\ref{fig:qed-processes}, can be studied 
in the multi-photon regime using the same Feynman diagrams, but with 
"dressed" (Dirac-Volkov) particle states and propagators \cite{VOLK35} 
because the particle is moving in the oscillating EM field.

\section{Gamma - electron scattering}
\subsection{Kinematics of the $e_i\,\gamma_i \to e_f\,\gamma_f$ scattering}
\label{sect:cinamatica-g-e-scatt}

\begin{figure}[h]
   \begin{minipage}{75mm}
\hspace{-10mm}
      \includegraphics[width=52mm]{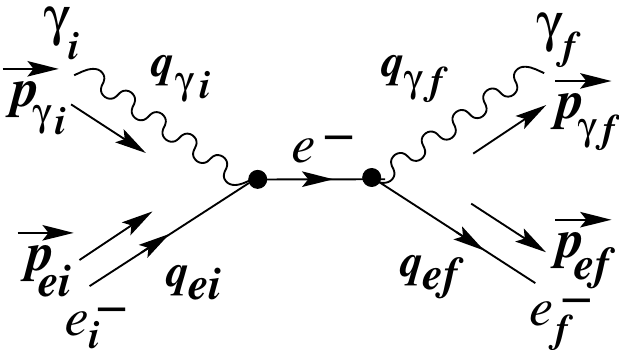}
   \end{minipage}
\\[12pt]
   \begin{minipage}{75mm}
\begin{tabular}{p{30mm}l}
$\epsilon_\gamma=h\nu$ 
& 
$\vec p_\gamma=\displaystyle{\frac{\epsilon_\gamma}{c}\,\,\vec n_\gamma}$
\\[6pt]
$\epsilon_e\!=\!\gamma_e\, m_e\,c^2$ 
& 
$\vec p_e\!=\!\displaystyle{\gamma_e\, m_e\,\vec v_e
\!=\!\frac{\epsilon_e}{c^2}\,\vec v_e\!=\!\frac{\epsilon_e}{c}\,\vec\beta_e}$
\\[6pt]
\multicolumn{2}{l}{
$\displaystyle{q\!\equiv\!\!\bigg(\frac{E}{c}\,,\,\vec p}\bigg)$
\qquad metric: \quad $g_{\mu\mu}\!=\!(1,-1,-1,-1)$
}
\\[12pt]
\multicolumn{2}{l}{
$\displaystyle{q^2=\frac{E^2}{c^2}-\vec p^{\,2} = m_e^2 c^2}$
}
\end{tabular}
   \end{minipage}
\vspace{1mm}
\hrule
\begin{tabular}{c|c|l}
& & \\
\hspace{-5mm}
$\displaystyle{q_{\gamma i}\!=\!\frac{\epsilon_{\gamma i}}{c}\bigg(1,\vec n_{\gamma i}\bigg)}$ 
&
$\displaystyle{q_{\gamma f}\!=\!\frac{\epsilon_{\gamma f}}{c}\bigg(1,\vec n_{\gamma f}\bigg)}$ 
& 
$\displaystyle{\beta_e\!=\!\frac{p_e\,c}{\epsilon_e}\!=\!\frac{v_e}{c}}$
\\[15pt]
\hspace{-5mm}
$\displaystyle{q_{ei}\!=\!\frac{\epsilon_{ei}}{c}\bigg(1,\vec\beta_{ei}\bigg)}$ 
&  
$\displaystyle{q_{ef}\!=\!\frac{\epsilon_{ef}}{c}\bigg(1,\vec\beta_{ef}\bigg)}$ 
& 
$\gamma_e\!=\!\displaystyle{\frac{\epsilon_e}{m_e\,c^2}\!=\!\frac{1}{\sqrt{1\!-\!\beta_e^2}}}$
\\ & &
\end{tabular}
\hrule
\caption{$e_i\,\gamma_i \to e_f\,\gamma_f$ scattering variables and kinematics} 
\label{fig:Cinamatica-Compton}
\end{figure}

The 4-momentum conservation, see Fig.\ref{fig:Cinamatica-Compton}:
   \begin{equation}\label{conserv-4-imp-Compton}
q_{\gamma i} + q_{ei} = q_{\gamma f} + q_{e f}
   \end{equation}
By squaring (\ref{conserv-4-imp-Compton}), with $q_\gamma^2=0$ and 
$q_e^2=m_e^2 c^2$, we have:
   \begin{equation}
\hspace{-15mm}
\label{prod-4-imp-Compton}
q_{\gamma i} \cdot q_{ei} = q_{\gamma f} \cdot q_{ef}
   \end{equation}
Multiply (\ref{conserv-4-imp-Compton}) by $q_{\gamma f}$ and 
using (\ref{prod-4-imp-Compton}) we have:
   \begin{equation}\label{rel-4-imp-Compton}
q_{\gamma f} \cdot \Big( q_{\gamma i} + q_{ei} \Big) = 
q_{\gamma i} \cdot q_{ei}
   \end{equation}

\begin{figure}[h]
      \includegraphics[width=55mm]{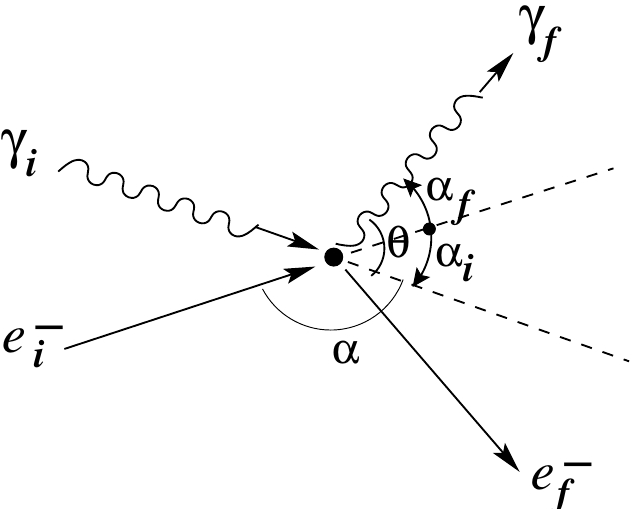}
\caption{Kinematics of the $\gamma_i\,e_i^- \to \gamma_f\,e_f^-$ scattering
}
\label{cinemat-e-g-scat}
\end{figure}

\begin{figure}[h]
      \includegraphics[width=55mm]{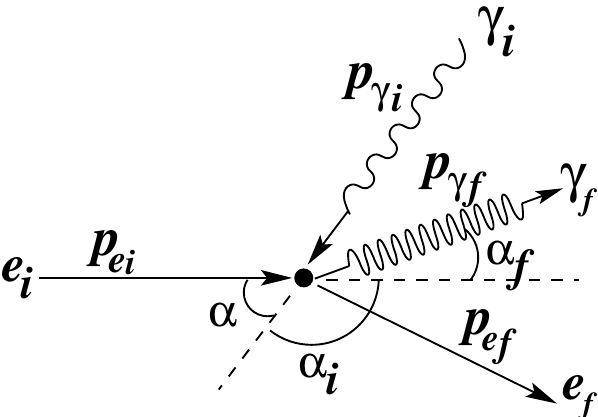}
\caption{The high energy $\gamma_f$-ray production geometry 
by inverse Compton scattering} 
\label{cinematica-SLAC}
\end{figure}

Expressing (\ref{rel-4-imp-Compton}) by 3-components, with the notations
of Fig.\ref{cinemat-e-g-scat}, we get the energy $\epsilon_{\gamma f}$ of the 
final photon as a function of the energy $\epsilon_{\gamma i}$ of the initial
photon for a given electron energy $\epsilon_{ei}$:
  \begin{equation}\label{energ-foton-Compton}
\epsilon_{\gamma f} =
\frac{\epsilon_{\gamma i}\, 
\epsilon_{ei}\,\Big(1-\beta_{ei} \cos\alpha_i\Big)}
{\epsilon_{\gamma i}\Big(1-\cos\theta\Big)
+\epsilon_{ei}\,\Big(1-\beta_{ei}\cos\alpha_f\Big)}
  \end{equation}

\noindent
In the initial electron rest frame:
$\beta_{ei}=0$\, ;\, $\gamma_{ei}=1\,  ;\, \epsilon_{ei}= m_e\,c^2$
\quad
the photon energy change is:
  \begin{equation}\label{modif-en-foton-sr-Compton}
\frac{\epsilon_{\gamma f}}{\epsilon_{\gamma i}}
=\frac{1}
{\displaystyle{
1+\frac{\epsilon_{\gamma i}}{m_e\,c^2}
\Big(1-\cos\theta\Big)}}
  \end{equation}

Multiply (\ref{rel-4-imp-Compton}) by $q_{ei}$, we have:
 \begin{equation}\label{rel-4-imp-Compt-bis}
q_{\gamma f}=
\frac{\Big(q_{\gamma i} \cdot q_{ei} \Big)\,\, q_{ei}}
{\Big(q_{\gamma i}\cdot q_{ei}\Big) + m_e^2 c^2}
 \end{equation}

The temporal component gives the energy dependence of the final photon on the 
initial electron energy:
\begin{equation}\label{energ-foton-final-Compton}
\epsilon_{\gamma f} = \frac{\Big(q_{\gamma i} \cdot q_{ei}\Big) \, \epsilon_{ei}}
{\Big(q_{\gamma i}\cdot q_{ei}\Big) + m_e^2 c^2}
\end{equation}

where (see Fig.\ref{cinematica-SLAC}) 
\begin{equation}\label{prod-qgi-qei}
q_{\gamma i} \cdot q_{ei}
=\frac{\epsilon_{\gamma i} \epsilon_{ei}}{c^2}\,
\Big(1\!+\!\beta_{ei}\cos\alpha\Big)
\end{equation}

That is, the final photon energy (\ref{energ-foton-final-Compton}) is
\begin{equation}\label{E-foton-final-Compton}
\epsilon_{\gamma f}
=\frac{\epsilon_{ei}}{
\displaystyle{
1+\frac{m_e^2\,c^4}{\epsilon_{\gamma i} \epsilon_{ei}}\, 
\frac{1}{\big(1+\beta_{ei}\cos\alpha \big)}}
}
\end{equation}

\subsection{The dressed electron mass in the electromagnetic field}

The motion of a free electron in an EM field can be described in terms of the 
interaction of the electron with a classical plane wave of frequency $\omega$. 
In general, such an electron will show an oscillatory motion with the 
same frequency $\omega$ and will radiate in turn. For a circularly polarized 
laser, EM waves with electric and magnetic components $E_L$ and $B_L$ have 
a constant amplitude and rotate with the angular frequency $\omega_L$ in a 
plane perpendicular to the direction of wave propagation. In this wave, 
the movement of the electron is circular with radius r, angular frequency 
$\omega_L$ and tangential speed $v_\perp=\omega_L\,r$ perpendicular to the 
direction of movement, parallel to the magnetic field vector $B_L$ \cite{SEIP17}. 

The circular motion of the electron is $m_e v_\perp^2/r=eE_L$ or
\begin{equation}\label{e-circ-motion}
p_\perp\omega_L=eE_L
\end{equation}
where $p_\perp\!=\!\gamma\, m_e v_\perp$ is the electron transverse momentum. 

For a relativistic electron we have  
$\beta_\perp\!\!=\!p_\perp c/\epsilon_e$ and $\gamma = \epsilon_e/m_e\,c^2$
and the transverse momentum will be:
\begin{equation}\label{mom-transv}
p_\perp = \beta_\perp \gamma\, m_e\, c
\end{equation}
The product: 
\begin{equation}\label{param-xi}
\xi = \beta_\perp\, \gamma
\end{equation}
defines the {\sl field strength parameter}.

A relativistic electron inside an EM plane wave field appears to have an 
"increased" mass. Indeed, based on the energy - mass relation of the 
electron:
\\

$\displaystyle{m_e^2 c^2 = \frac{\epsilon_e^2}{c^2} - \vec p^{\,2}}$
\quad \mbox{with}~ (\ref{mom-transv})~ and~ (\ref{param-xi})
\begin{equation}\label{el-energ-rel-mass}
m_e^2 c^2=\frac{\epsilon_e^2}{c^2} - p_\parallel^2 -\xi^2 m_e^2 c^2
\end{equation}
the relativistic energy of the electron in an EM field, now is:
\begin{equation}\label{relat-el-energ-in-EM}
\frac{\epsilon_e^2}{c^2} - p_\parallel^2 = m_e^2 c^2 \Big(1+\xi^2\Big)
\end{equation}
where $p_\parallel$ is the longitudinal momentum parallel to the direction of 
propagation of the wave. Heuristically, one can say that the electron 
behaves as if it had an effective mass \cite{KIBB65}:
\begin{equation}\label{el-eff-mass-EM}
\overline m_e = m_e \sqrt{1+\xi^2}
\end{equation}
This behavior is identifiable by a shift in the kinematic edge for 
Compton scattering, such that, the electron in an EM plane wave 
field is moving along the $p_\parallel$ and behaves as having a 
"dressed" mass $\overline m_e$.
 
Although this mass "increase" has been derived classically, the same 
relation (\ref{el-eff-mass-EM}) for the effective mass appears in the 
quantum treatment of the solutions of the Dirac equation for free 
electrons in the EM plane wave as Dirac-Volkov "dressed" states \cite{VOLK35}.

\subsection{Classical laser intensity parameter $\xi$ (nonlinearity 
charge-field coupling)}

The electric field strength parameter $\xi$ (\ref{param-xi}) can be 
expressed with the transverse momentum  (\ref{mom-transv}) or in 
connection (\ref{e-circ-motion}) with the electric field component $E_L$ 
or, as we will see, with the intensity $I_L$ of the
laser beam. For the moment it can be write:
\begin{equation}\label{redefined-xi}
\xi = \beta_\perp \gamma = \frac{e\,E_L}{m_e \omega_L c}
\end{equation}
The Lorentz factor $\gamma = 1/\sqrt{1-\beta_\perp^2}$  with (\ref{param-xi}) 
can be expressed as:  $\gamma = 1/\sqrt{1+\xi^2}$
and $\beta_\perp = \xi/\sqrt{1+\xi^2}$. 

Then the radius of electron's circular trajectory is less than the reduced 
laser wavelength $\lambdabar_L=\lambda_L/(2\pi)$:
\begin{equation}\label{raza-traiect-circ}
r=\frac{v_\perp}{\omega_L} = \frac{\beta_\perp c}{\omega_L}
=\frac{\xi}{\sqrt{1+\xi^2}}\,\frac{\lambda_L}{2\pi} \le \frac{\lambda_L}{2\pi}
\end{equation}

Now, it is convenient to redefine the parameter $\xi$ by squaring 
(\ref{redefined-xi}) as:
\begin{equation}\label{new-redefined-xi}
\xi^2 = \frac{e^2 \langle E_L^2\rangle}{m_e^2\,\omega_L^2\, c^2}
\end{equation}
where the average $\langle E_L^2\rangle$ is taken with respect to time. 

The $\xi^2$ (\ref{new-redefined-xi}) measures the average laser beam 
intensity $I_L$ expressed as usual in electro-dynamics by 
$\langle E_L^2\rangle$:
\begin{equation}\label{laser-intensity}
I_L=\epsilon_0\,c\,\langle E_L^2\rangle
\end{equation}
With the mean laser electric field as the r.m.s.
\vspace{1mm}

$\displaystyle{
E_L\approx \sqrt{\langle E_L^2\rangle} 
= \sqrt{\frac{1}{\epsilon_0\, c}}\,\sqrt{I_L}
}$
\qquad or 
\begin{equation}\label{laser-field-intensity}
E_L=1944\cdot\sqrt{I_L}
\qquad \mbox{for} \qquad I_L \mbox{~~in~~} W/cm^2
\end{equation}
Substituting $\langle E_L^2\rangle$ from (\ref{laser-field-intensity}) in 
(\ref{new-redefined-xi}) with $\omega_L=2\pi c/\lambda_L$, we get finally:
\begin{equation}\label{eval-xi}
\xi^2\!\!=\!3.65\times 10^{-19} I_L \lambda_L^2
 \,\,\,\, \mbox{for}\,\,
I_L \mbox{~in~} W/cm^2\,\, \mbox{and}\,\, \lambda_L \mbox{~in~} \mu m
\end{equation}
At ELI-NP for a laser wavelength $\lambda_L= 0.815~\mu m$ and the pulse 
intensity at the focus $I_L\sim 10^{22} W/cm^2$, we have $\xi\cong 50$.

\subsection{Physical interpretation of the laser intensity parameter $\xi$}

The laser intensity is connected with the energy transfer from the 
laser field to electron \cite{HEIN09}. The classical laser intensity parameter $\xi$
(\ref{redefined-xi}) can be interpreted with the work of the laser 
field over the electron Compton wavelength\, $\lambdabar_c$:
\begin{equation}\label{xi-interpret}
\xi=\frac{e\,E_L}{m_e\,c\,\omega_L}=e\,E_L\frac{\lambdabar_c}{\hbar\,\omega_L}
=e\,E_L\frac{\lambdabar_L}{m_e\,c^2}
\end{equation}

where \qquad 
$\displaystyle{
\lambdabar_c=\frac{\hbar}{m_e\,c} \quad    ;    \quad
 \lambdabar_L=\frac{c}{\omega_L}
}$
\vspace{3mm}

\begin{figure}[h]
      \includegraphics[width=50mm]{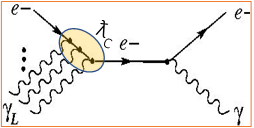}
\caption{Electron multi-photon laser interaction.} 
\label{multi-g-int}
\end{figure}

The product $e\,E_L\lambdabar_c$ represents the work of the laser field 
$E_L$ over the electron reduced Compton wavelength\, $\lambdabar_c$
(see Fig.\ref{multi-g-int}).
$\xi$ (\ref{xi-interpret}) is measuring this work in units of photon 
energy $\hbar\,\omega_L$. So $\xi$ gives the number of 
laser photons interacting along the $\lambdabar_c$.

$\xi$ is a classical parameter, because it is independent of $\hbar$ and
it measures the laser intensity as the number $n$ of $\gamma_L$ laser photons
interacting with the electron along the Compton wavelength\, $\lambdabar_c$.

Also, the last term of Eq. (\ref{xi-interpret}) can be interpreted as the
energy transfer to the electron over the laser wavelength\, $\lambdabar_L$
in units of electron rest mass $m_e\,c^2$.

If we consider the Schwinger electric field threshold $E_{cr}$ 
(\ref{eq:Schwingwr-field}), the $\xi$ parameter can be expressed as:
\begin{equation}\label{xi-new-interpret}
\xi=\frac{m_e\,c^2}{\hbar\,\omega_L}\,\frac{E_L}{E_{cr}}
=\frac{\lambdabar_L}{\lambdabar_c}\,\frac{E_L}{E_{cr}}
\end{equation}
The smallness of the factor $E_L/E_{cr}$ is compensated by the large 
ratio of laser to Compton wavelength $\lambdabar_L/\lambdabar_c$ of 
the order $10^6$ \cite{HEIN09}.

The $\xi$ parameter value encodes certain processes as:
\begin{itemize}
\item
$\xi \ll 1$ : the processes with minimum possible number of photons are the 
most probable. The probabilities equal the perturbation (linear) theory 
probabilities and plane waves play the role of individual photons. 
\item
$\xi \sim 1$ or $\xi > 1$ : the probabilities to absorb different number 
of photons become comparable and the process becomes multi-photon, i.e., 
the probability has an essentially non-perturbative (nonlinear) 
dependence on the field.
\item
$\xi\gg 1$ : the case of modern laser technology.
\end{itemize}

\subsection{Quantum nonlinearity parameter $\chi_e$}

As long as $\xi$ parameter (\ref{xi-interpret}) is given in relation to 
the photon field energy transfer over a reduced Compton wavelength  
$e\, E_L\, \lambdabar_c$ (in $\hbar\,\omega_L$ units), the quantum $\chi_e$ 
parameter \underline{is defined} in terms of the energy of the electron 
$\epsilon_e$ (in $m_e\,c^2$ units) and the laser field $E_L$ 
(in $E_{cr}$ units):
\begin{equation}\label{chi-param}
\chi_e=\frac{\epsilon_e}{m_e c^2} \frac{E_L}{E_{cr}}
=\gamma_e \frac{E_L}{E_{cr}}
=\gamma_e \frac{\hbar\,\omega_L}{m_e \,c^2} \,\xi
\end{equation}
here we used (\ref{xi-new-interpret}) to express connection with $\xi$.

If laser field is $E_L=E_{cr}$ and electron energy (\ref{work-el-mass}) 
$\epsilon_e=\epsilon_{cr}=e E_{cr} \lambdabar_c=m_e c^2$
is the work performed by the field $E_L$ 
over reduced Compton wavelength\, $\lambdabar_c$, then $\chi_e=1$.
This way $\chi_e$ compares the $\gamma_e E_L$ field strength in 
the rest frame of the electron, with the critical field $E_{cr}$ 
and measures the importance of quantum nonlinear effects in $e^+e^-$ 
vacuum pair production \cite{BERE82,BLAC18}.

In the context of SF-QED, the quantum nonlinearity parameter $\chi_e$ 
serves as a measure of the importance of nonlinear QED processes such as 
multi-photon Compton scattering, $e^+e^-$ pair production and 
photon-photon scattering. These processes become significant when 
the quantum nonlinearity parameter is of order unity or greater. 

\begin{table*}
\begin{minipage}{155mm}
\caption{The Feynman diagrams of some linear QED processes and the corresponding 
$\hat S$ matrix elements} 
\resizebox{\textwidth}{!}{
\begin{tabular}{||c|c|c||}
\hline \hline
Process
 & 
Feynman diagrams
 &
$\hat S$ matrix element
\\
\hline
   \begin{minipage}{40mm}
  \begin{center}
photon-electron scattering \\[3pt]
$\gamma+e^-\longrightarrow \gamma+e^-$
\\[12pt]
photon-positron scattering \\[3pt]
$\gamma+e^+\longrightarrow \gamma+e^+$
  \end{center}
   \end{minipage}
 &
   \begin{minipage}{30mm}
      \includegraphics[width=40mm,height=20mm]{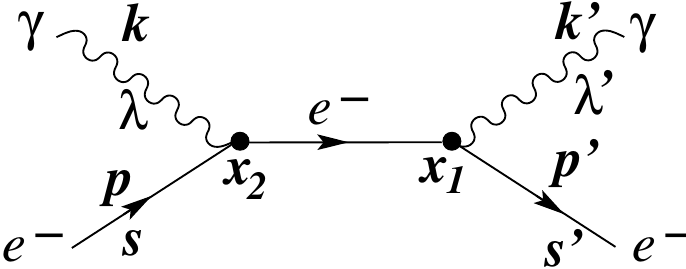}
   \end{minipage}
\hspace{15mm}
   \begin{minipage}{35mm}
\vspace{1mm}
      \includegraphics[width=35mm]{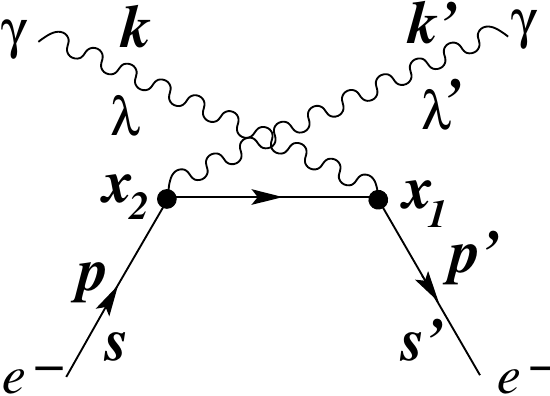}
\vspace{-2mm}
   \end{minipage}
 &
   \begin{minipage}{30mm}
\centering
$\big<\gamma, e^-\big| \hat S \big| \gamma, e^-\big>$
\\[12pt]
$\big<\gamma, e^+\big| \hat S \big| \gamma, e^+\big>$
   \end{minipage}
\\
\hline
   \begin{minipage}{40mm}
  \begin{center}
$e^+e^-$ pair anihilation \\[3pt]
$e^-+\,e^+\longrightarrow \gamma+\gamma$
  \end{center}
   \end{minipage}
 &
   \begin{minipage}{35mm}
\vspace{2mm}
      \includegraphics[width=25mm]{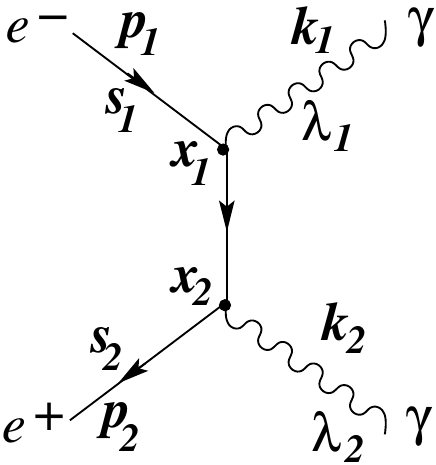}
\vspace{1mm}
   \end{minipage}
\hspace{5mm}
   \begin{minipage}{35mm}
      \includegraphics[width=30mm]{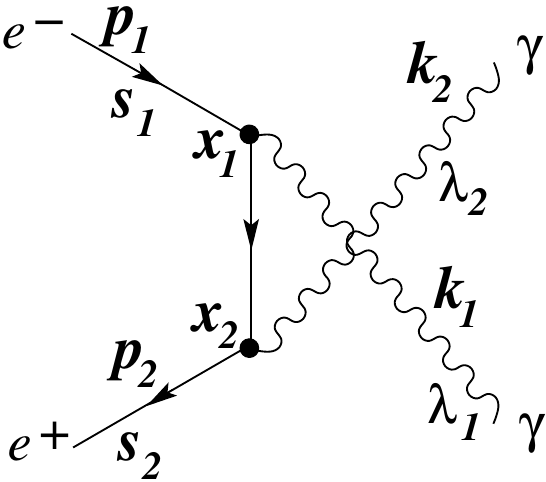}
   \end{minipage}
 &
$\big< \gamma, \gamma \big| \hat S \big| e^-, e^+ \big>$
\\
\hline
\begin{minipage}{40mm}
\begin{center}
$e^+e^-$ pair production \\[3pt]
$\gamma+\gamma \longrightarrow e^-+e^+$
  \end{center}
   \end{minipage}
 &
   \begin{minipage}{35mm}
\vspace{2mm}
      \includegraphics[width=25mm]{14-prod-pair-1.pdf}
\vspace{2mm}
   \end{minipage}
\hspace{5mm}
   \begin{minipage}{35mm}
      \includegraphics[width=30mm]{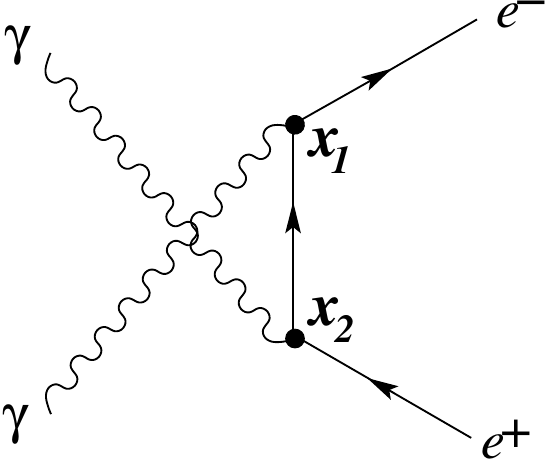}
   \end{minipage}
 &
$\big<e^-, e^+\big| \hat S \big| \gamma, \gamma\big>$
\\
\hline
   \begin{minipage}{40mm}
\begin{center}
$e^-$ M{\o}ller scattering \\[3pt]
$e^-+e^-\longrightarrow e^-+\,e^-$
\\[12pt]
$e^+$ M{\o}ller scattering \\[3pt]
$e^++e^+\longrightarrow e^++\,e^+$
\end{center}
   \end{minipage}
 &
   \begin{minipage}{35mm}
\vspace{2mm}
      \includegraphics[width=30mm,height=30mm]{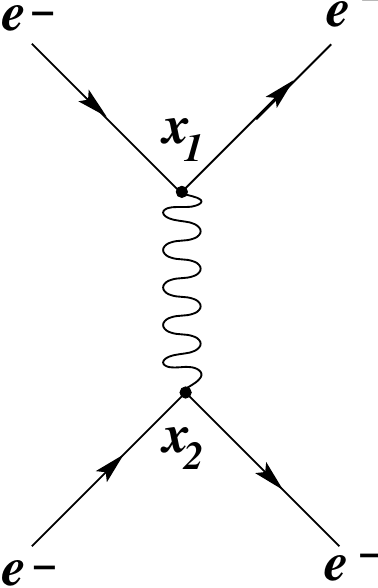}
\vspace{2mm}
   \end{minipage}
\hspace{5mm}
   \begin{minipage}{35mm}
\vspace{2mm}
      \includegraphics[width=30mm,height=30mm]{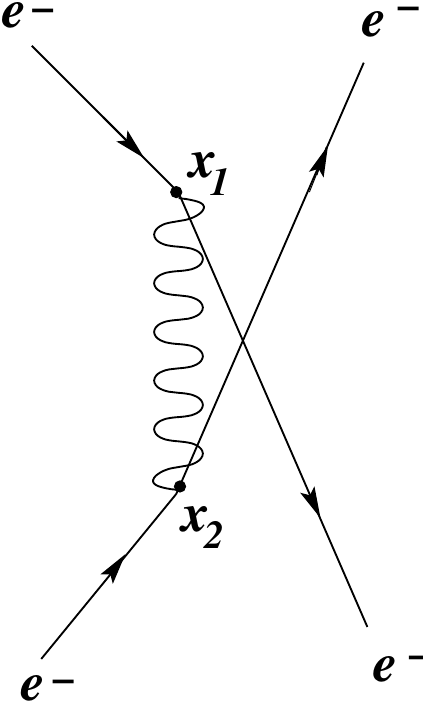}
\vspace{2mm}
   \end{minipage}
 &
   \begin{minipage}{30mm}
\centering
$\big<e^-, e^-\big| \hat S \big| e^-, e^-\big>$ 
\\[12pt]
$\big<e^+, e^+\big| \hat S \big| e^+, e^+\big>$ 
   \end{minipage}
\\
\hline
   \begin{minipage}{40mm}
\begin{center}
$e^-+e^+$ Bhabha scattering \\[3pt]
$e^-+e^+\longrightarrow e^-+e^+$
\end{center}
   \end{minipage}
 &
   \begin{minipage}{35mm}
      \includegraphics[width=40mm]{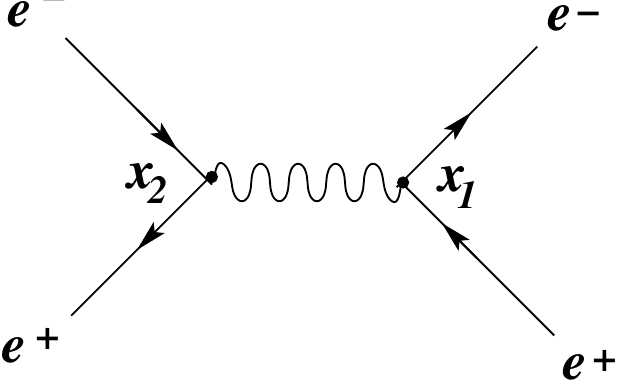}
   \end{minipage}
\hspace{10mm}
   \begin{minipage}{35mm}
\vspace{2mm}
      \includegraphics[width=30mm,height=30mm]{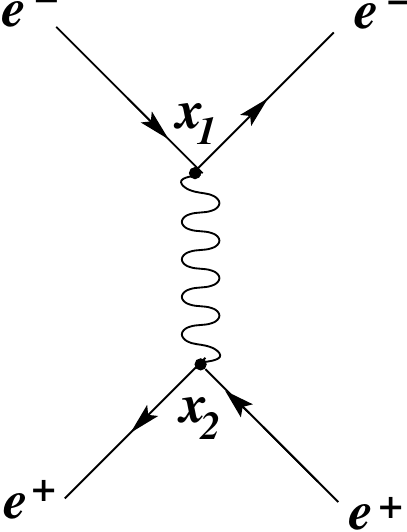}
\vspace{2mm}
   \end{minipage}
 &
$\big<e^-, e^+\big| \hat S \big| e^-, e^+\big>$ 
\\
\hline
   \begin{minipage}{40mm}
\begin{center}
Electron self energy \\[3pt]
$e^-\longrightarrow e^-$
\\[6pt]
Positron self energy \\[3pt]
$e^+\longrightarrow e^+$
\end{center}
   \end{minipage}
 &
   \begin{minipage}{60mm}
\vspace{2mm}
      \includegraphics[width=60mm]{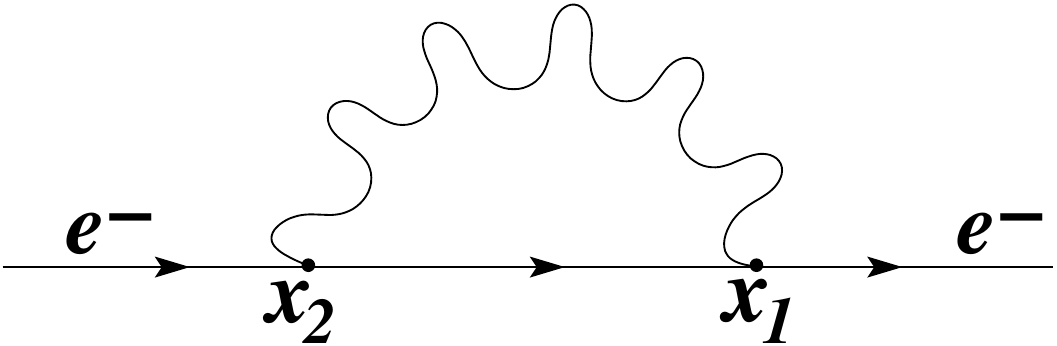}
\vspace{-2mm}
   \end{minipage}
 &
   \begin{minipage}{20mm}
$\big<e^-\big| \hat S \big| e^-\big>$ 
\\[12pt]
$\big<e^+\big| \hat S \big| e^+\big>$ 
   \end{minipage}
\\
\hline
   \begin{minipage}{40mm}
\begin{center}
Photon self energy \\[3pt]
$\gamma \longrightarrow \gamma$
\end{center}
   \end{minipage}
 &
   \begin{minipage}{65mm}
\vspace{1mm}
      \includegraphics[width=60mm]{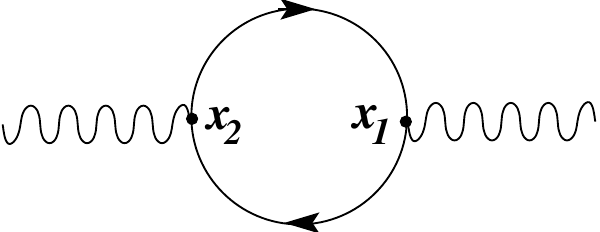}
\vspace{1mm}
   \end{minipage}
 &
$\big<\gamma \big| \hat S \big| \gamma\big>$ 
\\
\hline
   \begin{minipage}{40mm}
\begin{center}
Vacuum energy \\[3pt]
Vacuum $\longrightarrow$ Vacuum
\end{center}
   \end{minipage}
 &
   \begin{minipage}{45mm}
\vspace{1mm}
      \includegraphics[width=40mm]{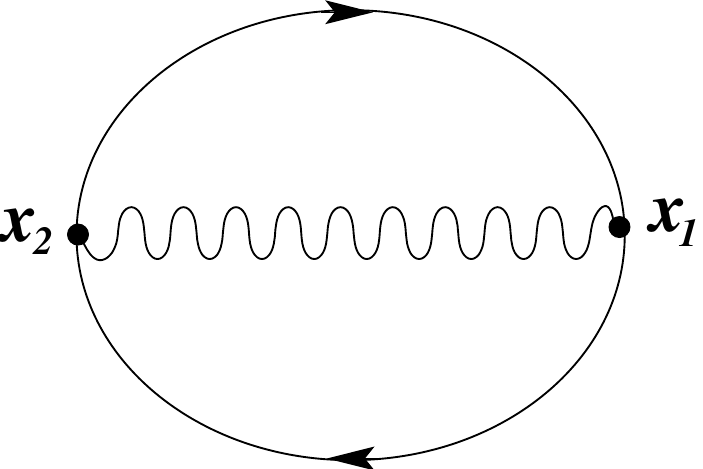}
\vspace{1mm}
   \end{minipage}
 &
$\big< 0 \big| \hat S \big|\,0 \big>$ 
\\
\hline
\hline
\end{tabular}
}
\label{tabl:proc-qed}
    \end{minipage}
\end{table*}

\section{Linear QED interaction processes}

The Feynman diagrams allows to determine the invariant amplitude of the 
QED processes, the $\hat S$ matrix elements and finally the cross section for 
the studied process. The Feynman diagrams of the interested linear QED processes are
shown in Table \ref{tabl:proc-qed}.

Evaluation of the Feynman diagrams uses the electromagnetic $\hat A_\mu(x)$ 
and Dirac $\hat\psi(x)$ and $\hat{\overline\psi}(x)$ field operators 
with the corresponding annihilation and creation components listed below: 
\small
\begin{equation}\label{dezv-Fourier-campuri}
\hspace{-2mm}
\left\{\hspace{-2mm}
\begin{array}{rcl}
\hat A_\mu(x) 
& \hspace{-2mm} = & \hspace{-2mm} 
\displaystyle{
\int\!\!\frac{d^3\vec k}{(2\pi)^3}\frac{1}{2\omega}
\Big[
\underbrace{
\hat a_\lambda(\vec k)
\epsilon_\mu^\lambda\,
e^{-i\,k\,\cdot\, x}}_
{\sim\, \hat A_\mu^-(x)} 
+\underbrace{
\hat a_\lambda ^\dagger(\vec k)\,
\epsilon_\mu^\lambda\,
e^{i\,k\,\cdot\,x}}_
{\sim\, \hat A_\mu^+(x)}
\Big]
}
\\[24pt]
\hat\psi(x) 
& \hspace{-2mm} = & \hspace{-2mm} 
\displaystyle{
\sum_s\! \int\!\!\frac{d^3\vec p}{(2\pi)^3}\frac{m}{\omega}
\Big[ 
\underbrace{
\hat b_s(\vec p)\,u_s(\vec p)\, 
e^{-i\,p\,\cdot\, x}}_
{\sim\, \hat\psi^-(x)}
+\underbrace{
\hat c_s^\dagger(\vec p)\,v_s(\vec p)\,
e^{i\,p\,\cdot\, x}}_
{\sim\, \hat\psi^+(x)}\Big]
}
\\[24pt]
\hat{\overline\psi}(x) 
& \hspace{-2mm} = & \hspace{-2mm} 
\displaystyle{
\sum_s\! \int\!\!\frac{d^3\vec p}{(2\pi)^3}\frac{m}{\omega}
\Big[ 
\underbrace{
\hat c_s(\vec p)\,\overline v_s(\vec p)\,
e^{-i\,p\,\cdot\, x}}_
{\sim\!\hat{\,\,\,\,\overline\psi{^-}}(x)} 
+ \underbrace{
\hat b_s^\dagger(\vec p)\,\overline u_s(\vec p)\,
e^{i\,p\,\cdot\, x}}_
{\sim\!\hat{\,\,\,\,\overline\psi{^+}}(x)} 
\Big]
}
\end{array}
\right.
\nonumber
\end{equation}
\normalsize

\noindent
where the field operators act with corresponding anihillation and 
creation operators:
\vspace{-4mm}

\begin{equation}\label{op-crea-anih}
\begin{tabular}{l}
\hline
\\[-9pt]
\hspace{5mm}
$\hat A_\mu^-(x) \to \hat a$ - photon anihillation in $x$
\\[6pt]
\hspace{5mm}
$\hat \psi^-(x) \to \hat b$ - electron anihillation in $x$
\\[6pt]
\hspace{5mm}
${\!\!\!\!\hat{\,\,\,\,\overline\psi{^-}}(x)}
 \to \hat c$ - positron anihillation in $x$
\\[3pt]
\hline
\\[-9pt]
\hspace{5mm}
$\hat A_\mu^+(x) \to \hat a^\dagger$ - photon creation in $x$
\\[6pt]
\hspace{5mm}
$\hat \psi^+(x) \to \hat c^\dagger$ - positron creation in $x$
\\[6pt]
\hspace{5mm}
${\!\!\!\!\hat{\,\,\,\,\overline\psi{^+}}(x)}
 \to \hat b^\dagger$ - electron creation in $x$
\\[3pt]
\hline
\end{tabular}
\end{equation}

\begin{figure}[ht]
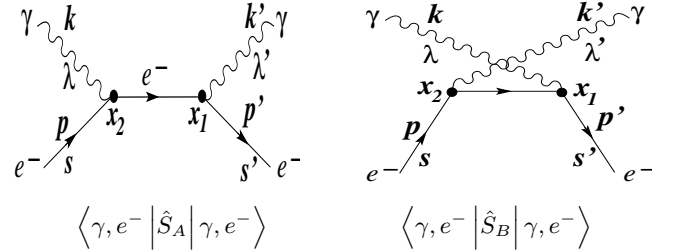

\begin{tabular}{cc}
     \includegraphics[width=38mm,height=24mm]{g-e-scat-1.pdf}
\hspace{3mm} & \quad
      \includegraphics[width=38mm,height=24mm]{g-e-scat-2.pdf}
\\[6pt]
$\Big<\gamma, e^-\,\Big|\hat S_A\Big|\,\gamma, e^-\Big>$
&
$\Big<\gamma, e^-\,\Big|\hat S_B\Big|\,\gamma, e^-\Big>$
\end{tabular}
\caption{
Feynman diagrams for the photon - electron (Compton) scattering
}
\label{fig:gamma-e-scatt}
\end{figure}

An example, for evaluation of the Feynman diagrams Fig.\ref{fig:gamma-e-scatt}
of the Compton (photon-electron) scattering, imply determination of the 
$\hat S$ matrix elements: 
$\big<\gamma , e^-\,\big|\,\hat S_A\, \big|\,\gamma , e^-\big>$ and
$\big<\gamma , e^-\,\big|\,\hat S_B\, \big|\,\gamma , e^-\big>$.

Using the appropriate creation/annihilation components, the scattering 
matrix for the two diagrams in Fig.\ref{fig:gamma-e-scatt}, are:
\begin{equation}
\label{S-A-explicit}
\displaystyle{
\hat S_A 
\!=\!-\left(\frac{e}{\hbar}\right)^{\!2}\!\!\int\!\! d^4x_1\,d^4x_2
\,N 
\Big(
\hspace{-6mm}
\stackrel{\hspace{-4mm}
\begin{minipage}{20mm}\centering crea.\\[-2pt] $e^-$ \\[-2pt] 
$\downarrow$\vspace{1.5mm}\end{minipage}}
{\!\!\!\!\hat{\,\,\,\,\overline\psi{^+}}\,}
\hspace{-12mm}
\stackrel{\hspace{-2mm}
\begin{minipage}{20mm}\centering crea.\\[-4pt] $\gamma$ \\ 
$\downarrow$\vspace{0.3mm}\end{minipage}}
{\not\!\!\hat A^+} 
\hspace{-7mm}
\contraction[2ex]{}
{\hat\psi}
{^-\Big)_{\!\!x_1}\!
\Big(}
{\!\!\!\!\hat{\,\,\,\,\overline\psi}}
\hat\psi^-
\Big)_{\!\!x_1}\!
\Big(
\!\!\!\!\hat{\,\,\,\,\overline\psi{^+}}
\hspace{-6mm}
\stackrel{\hspace{-3mm}
\begin{minipage}{20mm}\centering anih.\\[-2pt] $\gamma$ \\[-1pt] 
$\downarrow$\vspace{1.4mm}\end{minipage}}
{\!\!\not\!\!\hat A^-}
\hspace{-12mm}
\stackrel{\hspace{-2mm}
\begin{minipage}{20mm}\centering anih.\\[-4pt] $e^-$ \\[-2pt] 
$\!\!\!\downarrow$\vspace{0.1mm}\end{minipage}}
{\!\!\hat\psi^-}
\hspace{-9mm}\Big)_{\!\!x_2}
}
\nonumber
\end{equation}
\begin{equation}
\label{S-B-explicit}
\displaystyle{
\hat S_B 
\!=\!-\left(\frac{e}{\hbar}\right)^{\!2}\!\!\int\!\! d^4x_1\,d^4x_2
\,N 
\Big(
\hspace{-6mm}
\stackrel{\hspace{-4mm}
\begin{minipage}{20mm}\centering crea.\\[-2pt] $e^-$ \\[-2pt] 
$\downarrow$\vspace{1.5mm}\end{minipage}}
{\!\!\!\!\hat{\,\,\,\,\overline\psi{^+}}\,}
\hspace{-12mm}
\stackrel{\hspace{-2mm}
\begin{minipage}{20mm}\centering anih.\\[-4pt] $\gamma$ \\ 
$\downarrow$\vspace{0.3mm}\end{minipage}}
{\not\!\!\hat A^-} 
\hspace{-7mm}
\contraction[2ex]{}
{\hat\psi}
{^-\Big)_{\!\!x_1}\!
\Big(}
{\!\!\!\!\hat{\,\,\,\,\overline\psi}}
\hat\psi^-
\Big)_{\!\!x_1}\!
\Big(
\!\!\!\!\hat{\,\,\,\,\overline\psi{^+}}
\hspace{-6mm}
\stackrel{\hspace{-3mm}
\begin{minipage}{20mm}\centering crea.\\[-2pt] $\gamma$ \\[-1pt] 
$\downarrow$\vspace{1.4mm}\end{minipage}}
{\!\!\not\!\!\hat A^+}
\hspace{-12mm}
\stackrel{\hspace{-2mm}
\begin{minipage}{20mm}\centering anih.\\[-4pt] $e^-$ \\[-2pt] 
$\!\!\!\downarrow$\vspace{0.1mm}\end{minipage}}
{\!\!\hat\psi^-}
\hspace{-9mm}\Big)_{\!\!x_2}
}
\nonumber
\end{equation}
The above contractions are 
expressed by corresponding $S_F$ Feynman propagators between 
$x_2\to x_1$ (see Fig.\ref{fig:gamma-e-scatt}). 
\begin{center}
$
\contraction[2ex]{}
{\hat\psi}
{^-(x_1)}
{\!\!\!\!\hat{\,\,\,\,\overline\psi}}
\hat\psi^-(x_1)
\!\!\!\!\hat{\,\,\,\,\overline\psi{^+}}(x_2)
=i\,S_F(p+k)
$
\\[6pt]
$
\contraction[2ex]{}
{\hat\psi}
{^-(x_1)}
{\!\!\!\!\hat{\,\,\,\,\overline\psi}}
\hat\psi^-(x_1)
\!\!\!\!\hat{\,\,\,\,\overline\psi{^+}}(x_2)
=i\,S_F(p-k')
$
\end{center}

Individual invariant amplitude ${\cal M}_i$ is evaluated for 
each Feynman diagram relative to $\hat S_i$ matrix element. 
The $\hat S$ matrix element is given by total amplitude 
${\cal M}_{fi}$ and the phase space volume of the process:
\begin{equation}\label{total-scatt-matrix}
  \Big<\gamma , e^-\,
\Big|\,\hat S\, \Big|\,\gamma , e^-\Big>
= (2\pi)^4\,
\delta^4(p\!+\!k\!-\!p'\!-\!k')
\,{\cal M}_{fi}
\end{equation}
where the total amplitude ${\cal M}_{fi}$ is the sum of individual amplitudes:
\begin{equation}
\label{amplit-g-e-short-A+B}
\hspace{-2mm}
\left\{
\begin{array}{l}
{\cal M}_{fi}={\cal M}_{fi}^A + {\cal M}_{fi}^B
\\[12pt]
{\cal M}_{fi}^A
= \displaystyle{
\!-\!\left(\frac{e}{\hbar}\right)^{\!2}
}
\overline u_{s'}(\vec  p{\,'})\! \not\!\epsilon_{\lambda'}\, iS_F(p\!+\!k) 
\!\not\!\epsilon_\lambda\, u_s(\vec p)
\\[15pt]
{\cal M}_{fi}^B
= \displaystyle{
\!-\!\left(\frac{e}{\hbar}\right)^{\!2}
}
\overline u_{s'}(\vec  p{\,'})\! \not\!\epsilon_{\lambda'}\, iS_F(p\!-\!k') 
\!\not\!\epsilon_\lambda\, u_s(\vec p)
\end{array}
\right.
\end{equation}

This allows the determination of the cross section for 
$\gamma + e \to \gamma + e'$ scattering. 

Similarly, the cross sections for 
the other QED processes from Table \ref{tabl:proc-qed} could be evaluated.

\section{SF-QED interaction processes}

The SF-QED process can be understood as an interaction of an electron with 
a plane wave EM field of frequency $\omega$. In general, such an electron will 
exhibit oscillatory motion and in turn will radiate. 
Therefore, to calculate the  $\hat S$ matrix elements of the SF-QED 
processes, the technique of Feynman diagrams can be used, with the 
field states and propagators this time as dressed Dirac-Volkov state operators, 
with particles of effective mass ${\overline m}_e$ (\ref{el-eff-mass-EM}) 
and effective 4-momentum $q_\mu$ inside the field \cite{VOLK35}:
\begin{equation}\label{4-momentum-eff}
q_\mu=p_\mu + \frac{\xi^2 m^2}{1 (k\cdot p)}\,k_\mu
\end{equation}
where $k_\mu$ is the wave (laser) photon 4-momentum. 

The Feynman rules  summarized in \cite{SEIP17}, 
but used now for SF-QED processes,
with our notations (\ref{op-crea-anih}), are:
\begin{enumerate}
\item
External incoming or outgoing electrons 
are represented by laser dressed Volkov state operator 
$\hat \psi^-\!(x)$ (annihilation) or 
${\!\!\!\!\hat{\,\,\,\,\overline\psi{^+}}\!(x)}$ (creation),
respectively. For incoming and outgoing positrons, one uses the 
corresponding Volkov state operators 
${\!\!\!\!\hat{\,\,\,\,\overline\psi{^-}}(x)}$ (annihilation)
and $\hat \psi^+\!(x)$ (creation).
\item
An internal fermion line corresponds to the Dirac-Volkov propagator 
${\cal G}(x,y)$.
\item
Internal and external photon lines are translated into the free photon 
propagator and the free photon states, respectively, see e.g. 
\cite{PESK95}.
\item
Each interaction vertex corresponds to a factor $-i\,e\,\gamma^\mu$ and 
an integral $d^4x$.
\item
Symmetry factors for identical particles are the same as in usual QED.
\end{enumerate}

\subsection{The nonlinear inverse Compton scattering}
\label{sect:nonlin-C-s}

\begin{figure}[h]
\centering
      \includegraphics[width=50mm]{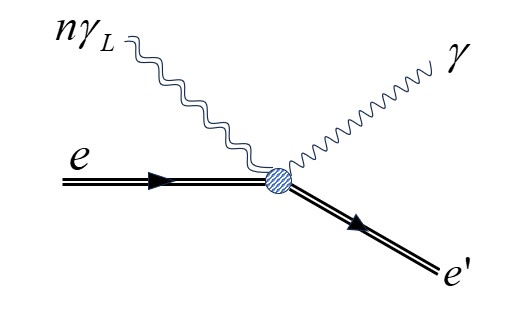}
\vspace{-2mm}
\caption{
The nonlinear Compton scattering. A dressed electron $e^-$ in a strong
EM field (double line) absorbs $n$ photons $\gamma_L$ (double sinusoids) 
and emits a high-energy photon $\gamma$ and recoils as $e'$.
}
\label{e-ng-double-line}
 \end{figure}

The first interaction process of the free electrons with strong EM 
fields we consider the nonlinear Compton scattering 
\cite{CHEN09,PIKE14,ROSE20,VOLK35,KIBB65,BERE82} in which 
an electron absorbs multiple photons and radiates a single 
high-energy photon (see Fig.\ref{e-ng-double-line}):
\begin{equation}\label{e-n-gamma}
e + n\,\gamma_L \rightarrow e' + \gamma
\end{equation}
\vspace{-5mm}

To study the SF-QED processes it is necessary to draw the corresponding 
Feynman diagrams with dressed particles to evaluate the invariant 
amplitude and finally to obtain the cross section.

In the experimental works with high power lasers, the inverse Compton 
scattering processes are used as the sources of the high energy photons 
(see Fig.\ref{fig:inv-Compt-scatt}).
\vspace{-3mm}

\subsection{Bremsstrahlung}
\label{sect:Bremsstrahhlung}

\begin{figure}
\hspace{-7mm}
\begin{tabular}{cc}
\includegraphics[width=40mm]{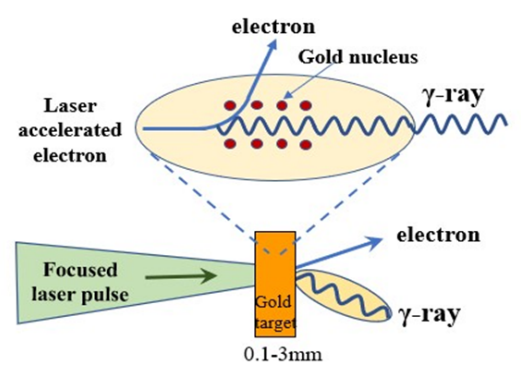}
&
\includegraphics[width=40mm]{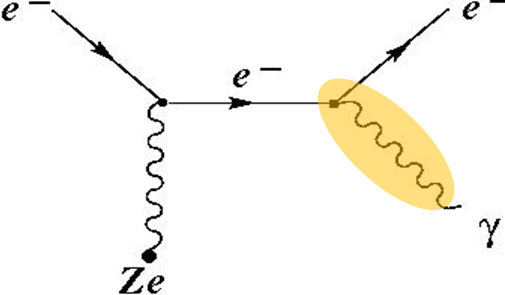}
\\
a) &  b)
\end{tabular}
\caption{
a) Experimental arrangement of the bremsstrahlung process; b) Feynman diagram of 
bremsstrahlung (the double line dressed electrons are not drawn).
} 
\label{fig:bremsstrahlung}
\end{figure}

Another process as a source of high energy gammas is bremsstrahlung 
interaction (see Fig.\ref{fig:bremsstrahlung}).

\begin{figure}[h]
\centering
      \includegraphics[width=55mm]{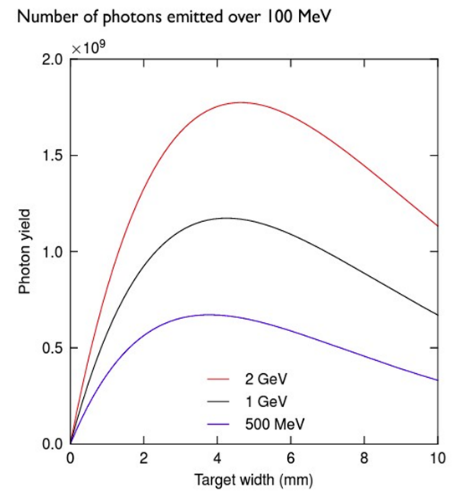}
\vspace{-2mm}
\caption{
High-energy photon distribution emitted from the gold target \cite{PIKE14a}
}
\label{fig:foton-en-distr-Au}
 \end{figure}

As an example, the ultra-relativistic electrons lose energy in a gold target 
almost solely by bremsstrahlung. The produced gamma rays in a particular 
arrangement \cite{PIKE14a} 
have the distributions presented in Fig.\ref{fig:foton-en-distr-Au}.
\vspace{-5mm}

\subsection{Breit-Wheeler pair production}

\begin{figure}[h]
\centering
\begin{tabular}{cc}
\includegraphics[width=27mm]{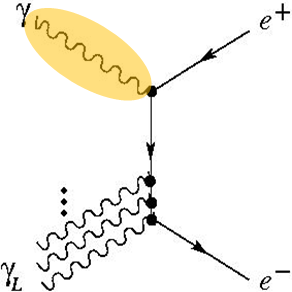}
& \qquad
\includegraphics[width=32mm]{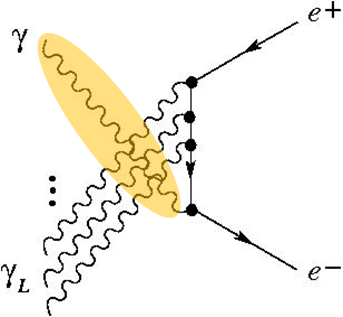}
\end{tabular}
\caption{
Feynman diagrams for $\gamma + n\,\gamma_L \rightarrow e^+ + e^-$  
Breit-Wheeler $e^+e^-$ pair production
(the double line dressed electrons are not drawn).
} 
\label{fig:BW-pair-prod}
\end{figure}

The high-energy gamma photons obtained from the previous processes of 
inverse Compton scattering or bremsstrahlung can interact with a 
multi-photon laser beam and as these photons propagate through the 
laser field can interact to produce electron-positron pairs:
\begin{equation}\label{g-n-gamma}
\gamma + n\,\gamma_L \rightarrow e^+ + e^-
\end{equation}
This is referred to as multi-photon Breit-Wheeler pair production and can 
be regarded as the materialization of a vacuum-polarization loop in a 
strong field. The cross section of the nonlinear Breit-Wheeler pair 
creation can be evaluated with the Feynman type diagrams  
Fig.\ref{fig:BW-pair-prod}, with dressed 
electrons (not drawn with double line in the figure).

\subsection{Bethe-Heitler pair production}

\begin{figure}[h]
\centering
\begin{tabular}{cc}
\includegraphics[width=27mm]{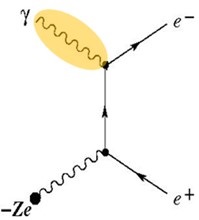}
& \qquad
\includegraphics[width=27mm]{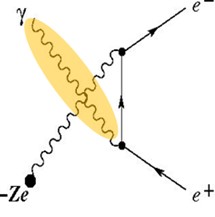}
\end{tabular}
\caption{
Feynman diagrams for Bethe-Heitler process
(the double line dressed electrons are not drawn).
} 
\label{fig:BH-proc}
\end{figure}

A process of $e^+e^-$ pair production possible to be stu\-died at 
ELI-NP is the interaction of the previously obtained high energy photons 
(by inverse Compton scattering or bremsstrahlung)
with virtual photons of a strong EM field of the atomic nucleus.
\begin{equation}\label{g-n-gamma-v}
\gamma + \gamma_V \rightarrow e^+ + e^-
\end{equation}
The evaluation of the cross section is done with the 
Feynman type diagrams Fig.\ref{fig:BH-proc}, but with the dressed electrons.

\begin{figure}[h]
\centering
\begin{tabular}{cc}
\includegraphics[width=30mm]{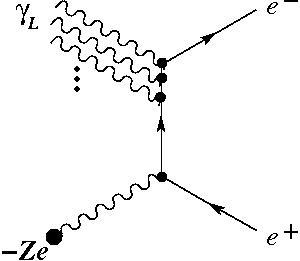}
& \qquad
\includegraphics[width=30mm]{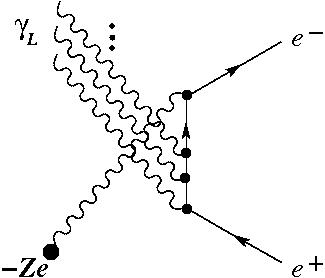}
\end{tabular}
\caption{
Feynman diagrams for multi-photon Bethe-Heitler $e^+e^-$ pair production
(the double line dressed electrons are not drawn).
} 
\label{fig:BH-gv-ng}
\end{figure}

Another possible Bethe-Heitler $e^+e^-$ pairs production is by 
multi-photon laser interaction with the virtual photons of an 
atomic nucleus field:
\begin{equation}\label{gv-n-gamma}
n\,\gamma + \gamma_V \rightarrow e^+ + e^-
\end{equation}
The evaluation of the cross section of this $e^+e^-$ pairs production 
in the field of the atomic nucleus is done with the Feynman type diagrams 
in Fig.\ref{fig:BH-gv-ng}.

\section{ELI-NP EXPERIMENTAL FACILITY}

The facilities at ELI-NP allow for the first time the use of the two 10 PW laser 
beams with intensities up to $10^{22}-10^{23} W/cm^2$ to study strong-field 
nonlinear QED interaction processes with laser beams \cite{TURC16,TURC19}.
The two 10 PW laser beams are extracted from the same laser pulse 
by splitting it into two pulses and are amplified on two identical 
amplifier chains. The two 10 PW pulses must be coherent. 
However, variations in the optical path traveled by the two pulses 
require additional adjustments to be made for "femtosecond" level 
synchronization.

\begin{figure}[h]
\hspace{8mm}
\includegraphics[width=35mm]{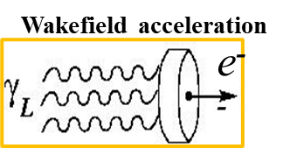}
\vspace{5mm}

\begin{minipage}{90mm}
\begin{center}
\includegraphics[width=85mm]{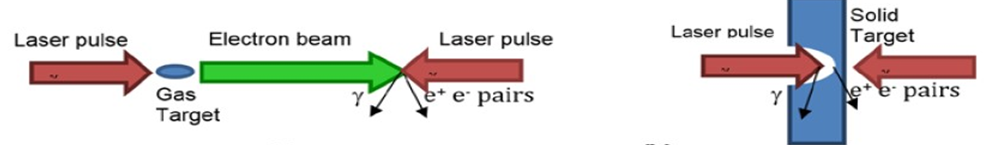}
\\
\hspace{10mm} a) \hspace{40mm} b)
\end{center}
\end{minipage}
\caption{
The two 10 PW laser pulses (red arrows) for SF-QED studies with gas (a) and solid (b) targets  
(I.C.E. Turcu et al. \cite{TURC16})}
\label{fig:two-lasers}
\end{figure}

There are two main types of experiments that could be performed 
(see Fig.\ref{fig:two-lasers}).

\begin{figure}[h]
\centering
\includegraphics[width=85mm]{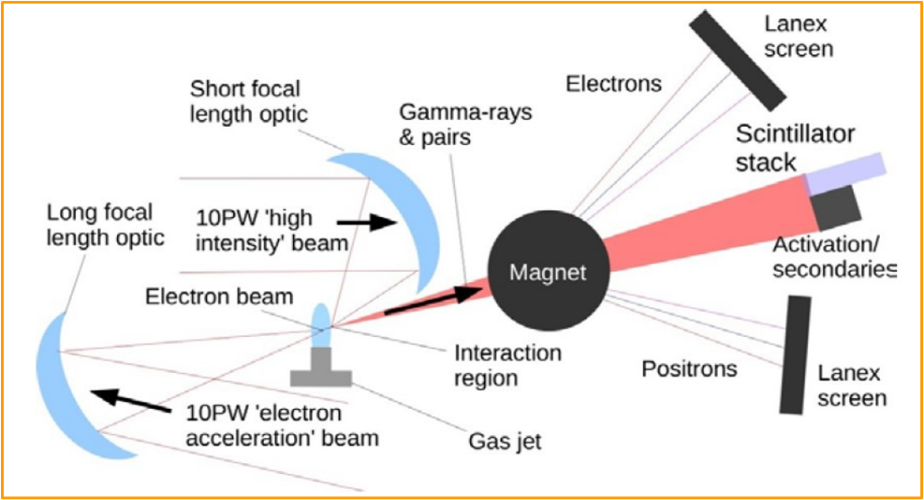}
\caption{
Layout of the gas target experimental components for the study of SF-QED 
processes (Reprinted from \cite{TURC16})}
\label{fig:SF-QED layout}
\end{figure}

\underline{In the first experiment}
 Fig.\ref{fig:two-lasers}.a), using one of the 10 PW laser beams as
the pump-laser focused with a large focal length mirror (F/20) on a 
gaseous target (gas jet or gas cell) which produces, by wakefield acceleration, 
relativistic electrons ($\gamma_e\gg 1$), see Fig.\ref{fig:SF-QED layout}. 
The second high-intensity 10 PW probe-laser 
beam is focused with an F/3 mirror onto the relativistic electron bunch,
Fig.\ref{fig:SF-QED layout}. 
Being exposed to the strongly focused laser field, they generate intense 
gamma rays and electron-positron pairs, through the nonlinear QED 
interaction processes that we intend to study.

The laser intensity of 10 PW focused in the focal spot with a diameter of 
5 $\mu$m is expected to be greater than $10^{22}~W/cm^2$. 
The pump and probe-laser are synchronized and delayed relative to each 
other as required. High-energy gamma photons can be measured with a 
gamma detector placed before the electron beam-dump. On the other hand, 
electron and positron spectra can be measured with dedicated spectrometers
Fig.\ref{fig:SF-QED layout}.

\underline{In the second experiment}
Fig.\ref{fig:two-lasers}.b), the 10 PW pump-laser is tightly 
focused by an F/3 mirror on a foil-solid target and produces relativistic 
electrons. The second laser, the 10 PW probe-laser is also tightly focused 
by an F/3 mirror on a solid target and then conveniently delayed relative 
to the pump-laser. The probe-laser produces a strong EM field to which 
the electrons are exposed. The solid target method is complementary to 
the gas target method. In the latter case, the number of relativistic 
electrons is very high, the target being a solid. This number is much 
higher than in the gas target or beam-beam method. On the other hand, 
the kinetic energy of the electrons is no longer as high as in the 
"beam-beam" method.

ELI-NP can prepare the E6 experimental area and the interaction chamber 
with the gas target, shown in Fig.\ref{fig:SF-QED layout}. 
The chamber contains the two pump \& probe colliding 10 PW 
laser beams. Measuring the characteristics of interacting SF-QED 
processes in intense laser fields is an experimental challenge: it 
requires high-sensitivity detectors for electrons and positrons in the 
presence of very high background levels of X-radiation and $\gamma$ photons.

The experimental area E6 with the two counter-propagating 10 PW laser beams 
focused on the gas and the solid target, can be prepared at ELI-NP.
The required equipment and configuration of the E6 experimental chamber with 
gas target are presented in Table \ref{tab:E6-equip}.

\begin{figure}[h]
\includegraphics[width=92mm]{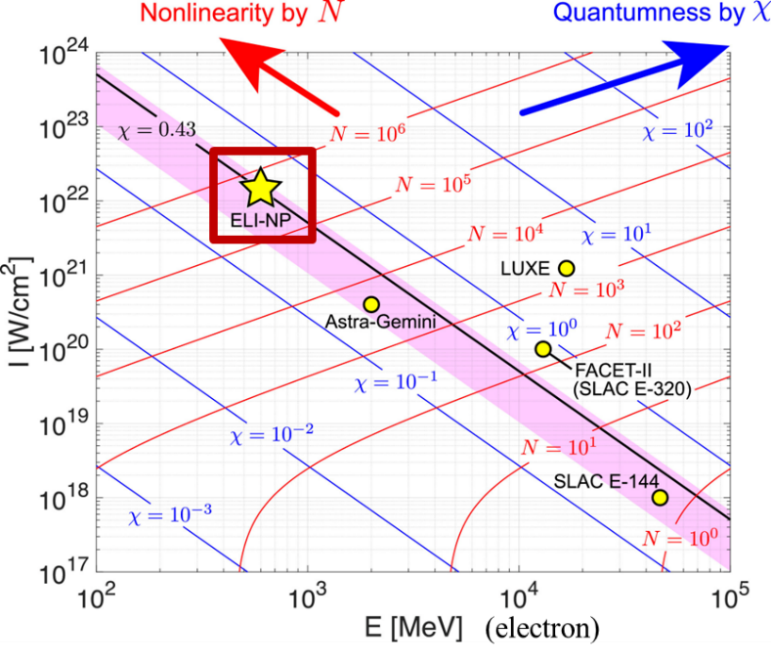}
\vspace{-3mm}
\caption{
The curves at given $N$ and $\chi$. $N$ is the number of absorbed laser
photon and $\chi$ is intensity parameter. The pink ribbon represents the
domain as $\chi \in [0.2, 0.5]$. We consider "linear" Compton scattering in
the area where $N\le 1$ (a single laser photon absorption). The star
symbol shows the parameter set at ELI-NP \cite{SETO21}.
}
\label{fig:ELI-NP-top}
\end{figure}

The possibilities presented by Keita Seto et al. \cite{SETO21} are shown 
in the diagram for the physical regime offered by ELI-NP 
(see Fig.\ref{fig:ELI-NP-top}). The high intensity of the ELI-NP laser 
allows obtaining a number of photons participating in an interaction 
$N >10^5$ (see the star in Fig.\ref{fig:ELI-NP-top}). 
The use of photons from 10 keV up to the GeV class should be considered.

\begin{table}
\caption{
E6 experimental chamber (laser-gas interaction)
}
\label{tab:E6-equip}
\begin{tabular}{|l|}
\hline 
\begin{minipage}{65mm}
\vspace{3mm}
\begin{center} \underline{E6 equipment consists of:} \end{center}
\begin{itemize}
\item
24 $m^3$ interaction chamber 
\item
ISO-7 clean room
\item
Opto-mechanical components
\item
Gas targets of various types
\item
30 m long focal, large aperture spherical mirror
\item
Laser and Plasma Diagnostics (1st or 2nd harmonic)
\item
Multi GeV electron spectrometers
\item
Gamma ray spectrometer
\end{itemize}
\vspace{-1mm}
\end{minipage}
\\ \hline

\begin{minipage}{85mm}
\vspace{2mm}
\begin{center}
\underline{E6 target area configuration (gas targets, QED)}
\end{center}
\begin{itemize}
\item
$2\times 10$ PW laser beams: 240 J, 23 fs, 810 nm, \\ 
$\sim$ 45 cm dia. FWHM 
\item
or 10 PW @ 1/60 Hz and 1 PW @ 1 Hz
\item
1 Short focal -\, parabolic mirrors F2.7
\item
1 Long focal $\sim$ 30 m -\, spherical mirror \\ 
$\sim$ F60 @ 10 PW ($\sim$ F160 @ 1 PW)
\item
1 Plasma mirror
\item
1 Cleanroom
\item
Experimental chamber: \\
$L\times W\times H$ of $4000\times 3300\times 1780~mm^3$
\end{itemize}
\vspace{-1mm}
\end{minipage}
\\
\hline 
\end{tabular}
\end{table}

\section{Ucoming Experiments}

Several petawatt-class lasers have been built worldwide. Currently, 
multi-PW and even 10 PW lasers have been built or are being planned 
to be built around the world \cite{DANS19}. Some of them have proposals 
to study fundamental processes in SF-QED regime in order to explore 
non-perturbative effects. These upcoming experiments include:

  - {\bf LUXE (Laser Und XFEL Experiment)} is a new experiment proposed 
at DESY and the European XFEL to study QED in the strong-field regime 
where QED becomes non-perturbative. 
It aims at studying high-field QED in electron-laser and photon-laser 
interactions, with the 16.5 GeV electron beam of the XFEL and a laser 
beam with power of up to 350 TW. The experiment will measure the 
spectra of electrons, positrons and photons \cite{LUXE19}.
\vspace{3mm}

 - {\bf ASTRA-GEMINI} (Central Laser Facility, STFC Rutherford Appleton 
Laboratory, Harwell Oxford, Dicot, UK) \cite{ROSE20}.
\vspace{3mm}

- {\bf E-320 experiment at FACET-II}, SLAC  
will collide 13 GeV electrons with ~10 TW laser pulses, to study 
fundamental strong-field QED processes \cite{FACE19}.
\vspace{3mm}

 - {\bf Apollon} in France. The ultra-intense and highly focused Apollo 
pulses make it possible to study the ultra-relativistic laser-plasma 
interaction regime, nonlinear Compton/Thomson scattering from 
laser-accelerated electron beams, the production of pairs in the 
presence of strong Coulombian fields \cite{APOL16}.

Other laser facilities with active SF-QED study programs include:
\vspace{3mm}

 - {\bf ZEUS facility at the University of Michigan}, once commissioned 
(late 2023), will use two laser pulses (with 2.5 PW and 0.5 PW), 
one to accelerate electrons in a laser wakefield accelerator 
(to either 10 GeV, or several GeV) and one to provide the EM field 
(with intensity $10^{21}~W/cm^2$, or $10^{23}~W/cm^2$), will allow exploration 
of fundamental yet unanswered questions regarding nonlinear quantum 
electrodynamics in relativistic plasmas, including quantum radiation 
reaction and electron-positron pair production mechanisms \cite{ZEUS22}.
\vspace{3mm}

 - {\bf SEL Station of Extreme Light (SEL)} facility in China, will be 
completed in 2025 and then open to users a 100-PW laser facility. 
It can provide focused intensity of more than $10^{23} W/cm^2$ \cite{SEL22}.
\vspace{2mm}

 - {\bf ELI-BL in Czech Republic} 
\vspace{-2mm}
{\begin{verbatim} https://www.eli-beams.eu/)\end{verbatim}
\vspace{2mm}

 - {\bf CALA in Germany}
\vspace{-2mm}
{\begin{verbatim} http://cala-laser.de/experiments/hf.html \end{verbatim}
\vspace{2mm}

 - {\bf J-Karen in Japan}, 
\vspace{-2mm}
{\begin{verbatim} https://doi.org/10.3390/qubs1010007 \end{verbatim}}
\vspace{2mm}

 - {\bf CORELS in Korea} 
\vspace{-2mm}
{\begin{verbatim} https://corels.ibs.re.kr/html/corels_en/ \end{verbatim}}

\section{CONCLUSIONS}

Here we presented the theoretical framework for the main SF-QED vacuum 
interaction processes possible to be studied at ELI-NP and the experimental 
possibilities offered by this laser infrastructure. 
The kinematics and dynamics of these processes were presented for
evaluation of the amplitudes necessary for cross section determination.

We analyzed the early experimental results as well as the new 
approaches of similar projects around the world. 
Based on these results, we can move on to the preparation, 
design and implementation of the experimental works to test 
some fundamental SF-QED interactions. For this purpose, 
it is necessary to go through some essential stages 
of the experimental studies at ELI-NP. 

It will be necessary to:
\begin{itemize}
\item
 evaluate the cross sections for the processes proposed to be studied
\item
 simulate the physical interaction processes 
\item
 preparation of the characteristic distributions on the available 
phase-space for these processes
\item
 detector system design to cover the available phase-space
\item
 the evaluation of the required statistics of the experimental data 
to achieve significant results to ve\-rify the QED predictions.
\item
 implementation of the detector system and performance of the 
experimental works.
\end{itemize}
Based on the results of this work, it is possible to proceed to the 
next stage of the design and realization of the experimental studies 
of the SF-QED processes at ELI-NP. 
\\[-12pt]

The proposed experimental configuration include:
\begin{itemize}
\item
 gas targets with the pump-laser beam focused by a long focal 
length (F/20 or F/80) mirror to drive a wakefield for electron 
bunch acceleration to multi-GeV energies and then exposed to the 
EM field of the probe-laser tightly focused (F/3).
\item
 solid targets with the pump and probe-laser beams focused on the target. 
\item
 vacuum QED experiments without any target but with similar 
interaction geometries and diagnostics to the ones above. 
\end{itemize}

  \begin{acknowledgments}
This work was supported by Institute of Atomic 
Physics and the Ministry of Research, Innovation and Digitalization 
under program ELI-NP-RO, Contract 8/ELI-RO (2020) - Romania.
  \end{acknowledgments}

\end{document}